\documentclass[aps,pra,amsmath,amssymb,preprint,letterpaper,superscriptaddress,11pt]{revtex4-1}

\usepackage{hyperref}
\usepackage{graphicx}
\usepackage{simplewick}

\newcommand{\trans}{\mathsf{T}}

\begin{document}

\title{$N$-electron Slater determinants from\\ non-unitary canonical
  transformations of fermion operators}

\author{Carlos A. Jim\'enez-Hoyos}
\affiliation{Department of Chemistry, Rice University, Houston, TX
  77005}

\author{R. Rodr\'iguez-Guzm\'an}
\affiliation{Department of Chemistry, Rice University, Houston, TX
  77005}
\affiliation{Department of Physics and Astronomy, Rice University,
  Houston, TX 77005}

\author{Gustavo E. Scuseria}
\affiliation{Department of Chemistry, Rice University, Houston, TX
  77005}
\affiliation{Department of Physics and Astronomy, Rice University,
  Houston, TX 77005}

\date{\today}

\begin{abstract}
Mean-field methods such as Hartree--Fock (HF) or
Hartree--Fock--Bogoliubov (HFB) constitute the building blocks upon
which more elaborate many-body theories are based on. The HF and HFB
wavefunctions are built out of independent quasi-particles resulting
from a unitary linear canonical transformation of the elementary
fermion operators. Here, we discuss the possibility of allowing the HF
transformation to become non-unitary. The properties of such HF vacua
are discussed, as well as the evaluation of matrix elements among such
states. We use a simple ansatz to demonstrate that a non-unitary
transformation brings additional flexibility that can be exploited in
variational approximations to many-fermion wavefunctions. The action
of projection operators on non-unitary based HF states is also
discussed and applied to the one-dimensional Hubbard model with
periodic boundary conditions.
\end{abstract}

\maketitle

\section{Introduction}

Mean-field methods such as Hartree--Fock (HF) or
Hartree--Fock--Bogoliubov (HFB) have become paradigmatic in the
description of many-fermion physics. These methods have found a wide
range of applications in nuclear structure theory, condensed matter
physics, and quantum chemistry. This is not only because they
constitute the simplest approximations to the exact many-body
wavefunction, but also because more elaborate correlated
approximations usually start from such independent quasi-particle
vacua (HF or HFB).

The HFB wavefunction developed to explain superconductivity relies on
the so-called Bogoliubov-Valatin \cite{bogoliubov1958,valatin1958}
transformation, which defines quasi-particle operators as linear
combinations of single-fermion creation and annihilation
operators. These are then used to form a quasi-particle product state,
the HFB wavefunction. Berezin \cite{berezin} studied the properties of
general linear transformations of fermionic operators within a
second-quantized framework. In this sense, one can consider the HF and
HFB wavefunctions as being built out of single quasi-particle
operators that result from a linear canonical transformation of the
elementary fermion ones.

A canonical transformation is understood in an algebraic framework as
that which preserves the Dirac bracket of the phase-space variables in
quantum mechanics (the position and momentum operators)
\cite{anderson1994}. In a second-quantized framework, this corresponds
to a transformation that preserves the anti-commutation rules of the
elementary fermion operators \cite{blaizot_ripka}. A linear canonical
transformation does not need to be unitary, although Dirac
\cite{dirac} and Weyl \cite{weyl} showed that unitary transformations
are canonical. Standard HF or HFB methods in several fields of
many-body physics are usually carried out using a unitary canonical
transformation. In this work, we study the possibility of constructing
$N$-particle Slater determinants resulting from non-unitary linear
canonical transformations. The extension to HFB determinants will be
discussed in a follow-up paper \cite{jimenez-hoyos2012b}.

We note that non-unitary canonical transformations have been discussed
in the literature before. They are discussed, for instance, by Blaizot
and Ripka \cite{blaizot_ripka} in the general context of canonical
transformations of second-quantized operators. They have been used by
Balian and Berezin \cite{balian1969} in the evaluation of matrix
elements between two different Bogoliubov states. Zhang and Tang
\cite{zhang1993}, and later Ma and Zhang \cite{ma1995}, have studied
the properties of linear canonical transformations of fermion
operators, including the non-unitary ones that we have just referred
to. We also mention the work of Anderson \cite{anderson1994}, where
the properties of non-unitary canonical transformations have been
discussed in a purely algebraic context, without reference to a
Hilbert space.

If a single Slater determinant is used as an ansatz for the
many-fermion wavefunction, the full flexibility that a non-unitary
canonical transformation affords is not evident because it does not
add additional degrees of freedom to those existing in a unitary
transformation. On the other hand, one can construct more general
ans\"atze that use the flexibility of such a non-unitary
transformation. We discuss here what may be the simplest,
two-determinant ansatz that exploits all the degrees of freedom that
define a non-unitary transformation for $N$-particle Slater
determinants. This idea has not been explored before in the
literature. We here derive all expressions required for the evaluation
of matrix elements between non-unitary based $N$-particle Slater
determinants. We also discuss the variational optimization of states
based on a non-unitary HF-type canonical transformation.

Our interest in non-unitary HF-type transformations originated from
our recent work on projected HF calculations for molecular systems
\cite{scuseria2011,jimenez-hoyos2012} and the two-dimensional Hubbard
Hamiltonian with periodic boundary conditions (PBC)
\cite{rodriguez-guzman2012}. The idea of using a symmetry-projected HF
state as an approximation to the many-body wavefunction was proposed
by L\"owdin \cite{lowdin1955} as early as 1955. We, building on
techniques developed and successfully applied in nuclear physics
\cite{ring_schuck,blaizot_ripka,sheikh2000,schmid2004,rodriguez-guzman2004,rodriguez-guzman2004b},
have shown that symmetry-projection out of the most general HF
transformation yields a multi-reference type wavefunction which can
account for a very significant part of the electron correlations. We
have observed that, the more general the transformation we use (or the
more symmetries that are broken), the better the resulting projected
wavefunction is able to account for the correlation structure of the
true Hamiltonian eigenvector. It is then natural to explore whether a
non-unitary canonical transformation, which has more degrees of
freedom than the unitary one commonly used, would yield additional
flexibility for HF wavefunctions in general, and projected HF states
in particular. This work describes our efforts along this line. We
show that, indeed, using a non-unitary canonical transformation, one
can build more flexible ans\"atze (based on $N$-particle Slater
determinants) from which additional correlations can be accounted for
in variational approximations.

This paper is organized as follows. In section \ref{sec:can}, we
discuss some general properties of linear canonical transformations of
fermion operators. We proceed to show in section \ref{sec:det} how to
construct $N$-particle Slater determinants based on such
transformations. Section \ref{sec:Thou} discusses our extension of
Thouless' theorem for non-unitary Slater determinants. This is
followed by section \ref{sec:matel}, where we use this theorem to
derive the form of matrix elements between non-unitary $N$-particle
Slater determinants. In section \ref{sec:ansatz}, we introduce a
two-determinant ansatz that displays the full flexibility of a
non-unitary transformation. We show in section \ref{sec:proj_ansatz}
how such an ansatz can be used in projected HF approaches. This is
followed by an illustrative application of the proposed wavefunction
ans\"atze to the one-dimensional Hubbard Hamiltonian with PBC in
section \ref{sec:hubbard}.

\section{Canonical transformations \label{sec:can}}

We start by introducing a set of fermion annihilation and creation
operators $\boldsymbol{c} = \{ c_k, c_k^\dagger \}$, which obey the
standard anti-commutation relations
\begin{center}
  \begin{tabular}{c @{\hspace{1cm}} c @{\hspace{1cm}} c}
    $\Big[ c_k, c_j \Big]_+ = 0$, &
    $\Big[ c_k^\dagger, c_j^\dagger \Big]_+ = 0$, &
    $\Big[ c_k, c_j^\dagger \Big]_+ = \langle k|j \rangle =
      \delta_{jk}$,
  \end{tabular}
\end{center}
where $|k \rangle$ ($\langle k|$) is a single-particle ket (bra)
state.

We now introduce a new set of fermion operators $\boldsymbol{\beta} =
\{ \beta_k, \bar{\beta}_k^\dagger \}$, which is related to the
original one by the linear transformation
\begin{equation}
  \begin{pmatrix} \beta \\ \bar{\beta}^\dagger \end{pmatrix} =
  \begin{pmatrix} U^\dagger & V^\dagger \\ Y^\trans &
    X^\trans \end{pmatrix}
  \begin{pmatrix} c \\ c^\dagger \end{pmatrix},
  \label{Tcan}
\end{equation}
where we have arranged the sets of fermion operators $\boldsymbol{c}$
and $\boldsymbol{\beta}$ into single columns. Here, $U$, $V$, $Y$, and
$X$ are arbitrary $M \times M$ matrices, where $M$ is the dimension of
the single-particle space. For compactness, we write the
transformation defined by Eq. \ref{Tcan} as
\begin{equation}
  \boldsymbol{\beta} = T \, \boldsymbol{c}.
\end{equation}

It should be stressed that we have not enforced the relation
$\bar{\beta}^\dagger = (\beta)^\dagger$ in Eq. \ref{Tcan} as this
leads to a standard unitary transformation. One can show
\cite{blaizot_ripka} that the transformation is unitary if the matrix
$T$ satifies
\begin{equation}
  T^\ast = \sigma \, T \, \sigma,
  \label{tunitary}
\end{equation}
where the matrix $\sigma$ is given by
\begin{equation}
  \sigma = \begin{pmatrix} 0 & 1 \\ 1 & 0 \end{pmatrix}.
\end{equation}
Here, Eq. \ref{tunitary} implies $U = X$ and $V = Y$.

We do insist, on the other hand, in making our transformation
canonical, which implies preserving the appropriate anti-commutation
relations, that is,
\begin{center}
  \begin{tabular}{c @{\hspace{1cm}} c @{\hspace{1cm}} c}
    $\Big[ \beta_k, \beta_j \Big]_+ = 0$, &
    $\Big[ \bar{\beta}_k^\dagger, \bar{\beta}_j^\dagger \Big]_+ = 0$, &
    $\Big[ \beta_k, \bar{\beta}_j^\dagger \Big]_+ = \delta_{jk}$.
  \end{tabular}
\end{center}

It is not difficult to prove \cite{blaizot_ripka} that the
transformation $T$ is canonical if it obeys
\begin{equation}
  T \, \sigma \, T^\trans = \sigma,
  \label{cancond}
\end{equation}

Using Eq. \ref{cancond}, one can easily deduce the form of the inverse
transformation
\begin{equation}
  T^{-1} = \begin{pmatrix} X & V^\ast \\ Y & U^\ast \end{pmatrix}.
  \label{Tinv}
\end{equation}
Equation \ref{cancond} also provides the conditions that the matrices
$U$, $V$, $X$, and $Y$ must satisfy for $T$ to define a canonical
transformation. Those are given by
\begin{subequations}
  \begin{align}
    U^\dagger \, X + V^\dagger \, Y &= 1, \\
    X^\trans \, U^\ast + Y^\trans \, V^\ast &=1, \\
    U^\dagger \, V^\ast + V^\dagger \, U^\ast &=0, \\
    Y^\trans \, X + X^\trans \, Y &= 0.
  \end{align}
\end{subequations}
Note that the matrices $U^\dagger \, V^\ast$ and $Y^\trans \, X$ are
anti-symmetric.

The matrices $T$ form a group (the fermion group described by Ma and
Zhang \cite{ma1995}) isomorphic to the group of orthogonal matrices of
dimension $2 M$ [$O(2M,C)$] \cite{blaizot_ripka}. On the other hand,
the set of matrices $T$ for which the transformation is unitary form a
group isomorphic to the group of real orthogonal matrices of dimension
$2M$ [$O(2M)$]. There are twice as many degrees of freedom in choosing
a general non-unitary transformation than in a unitary one.

We close this section by noting that the transformation defined in
Eq. \ref{Tcan} is more naturally understood as a linear transformation
if one introduces an operator $S$ such that
\begin{equation}
  \boldsymbol{\beta} = S \, \boldsymbol{c} \, S^{-1} =
  T \, \boldsymbol{c}.
\end{equation}
The form of the operator $S$ has been discussed by Blaizot and Ripka
\cite{blaizot_ripka}, Zhang and Tang \cite{zhang1993}, and Ma and
Zhang \cite{ma1995}.

\section{$N$-electron Slater determinants \label{sec:det}}

In this section, we discuss the construction of $N$-particle Slater
determinants using quasi-particle operators resulting from canonical
transformations of the elementary fermion ones. This is discussed in
detail by Navon \cite{navon1969}, as well as in several textbooks in
many-body physics.

In standard ({\em i.e.} unitary) HF theory, an $N$-electron Slater
determinant is constructed out of a set $N$ hole creation ($\{
b_h^\dagger \}$) and $M-N$ particle annihilation ($\{ b_p \}$)
operators, each of them resulting from a linear combination of the
elementary operators $\{ c_k, c_k^\dagger \}$:
\begin{subequations}
  \label{operator1}
  \begin{align}
    b_h^\dagger &= \sum_j D_{jh}^\ast \, c_j^\dagger, \\
    b_p &= \sum_j D_{jp} \, c_j.
  \end{align}
\end{subequations}
Using standard notation, the first $N$ columns in $D$ (which we write
as $D_h$) represent the hole states, while the last $M-N$ columns
(which we write as $D_p$) represent the particle states.

The transformation from the elementary operators to the set of HF
operators constructed above can be written as
\begin{equation}
  \begin{pmatrix} b_h^\dagger \\ b_p \\
    b_h \\ b_p^\dagger \end{pmatrix} =
  \begin{pmatrix}
    \mathbf{0}_{N \times M} & D_h^\dagger \\
    D_p^\trans & \mathbf{0}_{(M-N) \times M} \\
    D_h^\trans & \mathbf{0}_{N \times M} \\
    \mathbf{0}_{(M-N) \times M} & D_p^\dagger
  \end{pmatrix}
  \begin{pmatrix} c \\ c^\dagger \end{pmatrix},
\end{equation}
where we have implicitly assumed the transformation to be unitary.

The above transformation is canonical if the HF operators satisfy the
(non-trivial) anti-commutation relations
\begin{center}
  \begin{tabular}{c @{\hspace{1cm}} c @{\hspace{1cm}} c}
    $\Big[ b_h, b_{h'}^\dagger \Big]_+ = \delta_{h'h}$, &
    $\Big[ b_p, b_{p'}^\dagger \Big]_+ = \delta_{p'p}$, &
    $\Big[ b_p, b_{h}^\dagger \Big]_+ = 0$.
  \end{tabular}
\end{center}
These conditions restrict the form of the matrix $D$ according to
\begin{subequations}
  \begin{align}
    \Big[ b_h, b_{h'}^\dagger \Big]_+ &=
    \sum_{jk} D_{jh} \, D_{kh'}^\ast \, \delta_{jk} =
    \left( D^\dagger \, D \right)_{h'h} = \delta_{h'h}, \\
    \Big[ b_p, b_{p'}^\dagger \Big]_+ &=
    \sum_{jk} D_{jp} \, D_{kp'}^\ast \, \delta_{jk} =
    \left( D^\dagger \, D \right)_{p'p} = \delta_{p'p}, \\
    \Big[ b_p, b_{h}^\dagger \Big]_+ &=
    \sum_{jk} D_{jp} \, D_{kh}^\ast \, \delta_{jk} =
    \left( D^\dagger \, D \right)_{hp} = 0.
  \end{align}
\end{subequations}
The first equation implies orthonormality of the hole states, the
second one orthonormality of the particle states, and the last one
corresponds to orthogonality between hole and particle states. All
these conditions are summarized in the requirement $D^\dagger \, D =
\mathbf{1}$.

One could allow the HF transformation described previously to become
non-unitary by introducing, in addition to the operators described by
Eq. \ref{operator1}, another set of hole and particle operators, $\{
\bar{b}_h, \bar{b}_p^\dagger \}$, given by
\begin{subequations}
  \label{operator2}
  \begin{align}
    \bar{b}_h &= \displaystyle \sum_j \bar{D}_{jh} \, c_j, \\
    \bar{b}_p^\dagger &= \displaystyle \sum_j \bar{D}_{jp}^\ast \,
      c_j^\dagger,
  \end{align}
\end{subequations}

A non-unitary transformation can then be built as
\begin{equation}
  \begin{pmatrix} b_h^\dagger \\ b_p \\
    \bar{b}_h \\ \bar{b}_p^\dagger \end{pmatrix} =
  \begin{pmatrix}
    \mathbf{0}_{N \times M} & D_h^\dagger \\
    D_p^\trans & \mathbf{0}_{(M-N) \times M} \\
    \bar{D}_h^\trans & \mathbf{0}_{N \times M} \\
    \mathbf{0}_{(M-N) \times M} & \bar{D}_p^\dagger
  \end{pmatrix}
  \begin{pmatrix} c \\ c^\dagger \end{pmatrix}.
  \label{HFtrans}
\end{equation}
It is a canonical transformation if the (non-trivial) anti-commutation
relations
\begin{center}
  \begin{tabular}{c @{\hspace{1cm}} c @{\hspace{1cm}} c @{\hspace{1cm}} c}
  $\Big[ b_p, b_h^\dagger \Big]_+ = 0$, &
  $\Big[ \bar{b}_h, \bar{b}_p^\dagger \Big]_+ = 0$, &
  $\Big[ \bar{b}_h, b_{h'}^\dagger \Big]_+ = \delta_{h'h}$, &
  $\Big[ b_p, \bar{b}_{p'}^\dagger \Big]_+ = \delta_{p'p}$,
  \end{tabular}
\end{center}
are satisfied. These conditions restrict the form of the matrices $D$
and $\bar{D}$ according to
\begin{subequations}
  \label{HFanticom}
  \begin{align}
    \Big[ b_p, b_{h}^\dagger \Big]_+ &=
    \sum_{jk} D_{jp} \, D_{kh}^\ast \, \delta_{jk} =
    \left( D^\dagger \, D \right)_{hp} = 0, \\
    \Big[ \bar{b}_h, \bar{b}_{p}^\dagger \Big]_+ &=
    \sum_{jk} \bar{D}_{jh} \, \bar{D}_{kp}^\ast \, \delta_{jk} =
    \left( \bar{D}^\dagger \, \bar{D} \right)_{ph} = 0, \\
    \Big[ \bar{b}_h, b_{h'}^\dagger \Big]_+ &=
    \sum_{jk} \bar{D}_{jh} \, D_{kh'}^\ast \, \delta_{jk} =
    \left( D^\dagger \, \bar{D} \right)_{h'h} = \delta_{h'h}, \\
    \Big[ b_p, \bar{b}_{p'}^\dagger \Big]_+ &=
    \sum_{jk} D_{jp} \, \bar{D}_{kp'}^\ast \, \delta_{jk} =
    \left( \bar{D}^\dagger \, D \right)_{p'p} = \delta_{p'p}.
  \end{align}
\end{subequations}
The first two equations imply orthogonality of the hole and particle
states in $D$ and $\bar{D}$. The last two equations imply a
bi-orthonormality between the hole and particle orbitals in $D$ and
$\bar{D}$. Note that the last two conditions are satisfied by choosing
$\bar{D}^\dagger = D^{-1}$, but the orthogonality among hole and
particle states has to be separately imposed.

Let us remark that, if the HF operators $\{ b_h^\dagger, b_p,
\bar{b}_h, \bar{b}_p^\dagger \}$ define a canonical transformation,
the inverse transformation is given by (see Eq. \ref{Tinv})
\begin{equation}
  \begin{pmatrix} c \\ c^\dagger \end{pmatrix} =
  \begin{pmatrix}
    \mathbf{0}_{M \times N} & \bar{D}_p^\ast &
      D_h^\ast & \mathbf{0}_{M \times (M-N)} \\
    \bar{D}_h & \mathbf{0}_{M \times (M-N)} &
      \mathbf{0}_{M \times N} & D_p
  \end{pmatrix}
  \begin{pmatrix} b_h^\dagger \\ b_p \\ \bar{b}_h
    \\ \bar{b}_p^\dagger \end{pmatrix}.
  \label{HFinv}
\end{equation}

The bi-orthonormal Slater determinants $|\Phi \rangle$ and
$|\overline{\Phi} \rangle$ are produced when the set of operators $\{
b_h^\dagger, \bar{b}_h^\dagger \}$ act on the bare fermion vacuum $|-
\rangle$, {\em i.e.},
\begin{align}
  |\Phi \rangle = \prod_h b_h^\dagger |- \rangle, \\
  |\overline{\Phi} \rangle = \prod_h \bar{b}_h^\dagger |- \rangle.
\end{align}
They satisfy the bi-orthonormality condition $\langle \overline{\Phi}
| \Phi \rangle = 1$.

One can easily show that $|\Phi \rangle$ and $|\overline{\Phi}
\rangle$ act as vacua to a certain set of hole or particle states:
\begin{center}
  \begin{tabular}{c @{\hspace{1cm}} c}
    $b_h^\dagger |\Phi \rangle = 0 \quad \forall \quad b_h^\dagger$, &
    $b_p |\Phi \rangle = 0 \quad \forall \quad b_p$, \\
    $\bar{b}_h^\dagger |\overline{\Phi} \rangle = 0 \quad \forall
      \quad \bar{b}_h^\dagger$, &
    $\bar{b}_p |\overline{\Phi} \rangle = 0 \quad \forall \quad
      \bar{b}_p$.
  \end{tabular}
\end{center}

\section{Thouless' theorem for $N$-electron Slater determinants \label{sec:Thou}}

In standard ({\em i.e.} unitary) HF, there is a theorem due to
Thouless \cite{thouless1960} which reads:

\begin{quote}
  {\bf Theorem.} Given a Slater determinant $|\Phi_0 \rangle$ which is
  a vacuum to the operators $\{ b_h^\dagger, b_p \}$, any $N$-particle
  Slater determinant $|\Phi_1 \rangle$ which is not orthogonal to
  $|\Phi_0 \rangle$ can be written in the form
  \begin{equation}
    |\Phi_1 \rangle =
    \mathcal{N} \, \exp \left( \sum_{ph} Z_{ph} \, b_p^\dagger \, b_h
      \right) |\Phi_0 \rangle,
    \label{Thou1}
  \end{equation}
  where $\mathcal{N} = \langle \Phi_0 | \Phi_1 \rangle$ is a
  normalization constant and the coefficients $Z_{ph}$ are uniquely
  determined. Conversely, any wavefunction of the form of
  Eq. \ref{Thou1}, where $|\Phi_0 \rangle$ is a Slater determinant, is
  also an $N$-particle Slater determinant.
\end{quote}

For Slater determinants built out of operators resulting from a
non-unitary linear canonical transformation, the equivalent theorem
reads

\begin{quote}
  {\bf Theorem.} Given a Slater determinant $|\Phi_0 \rangle$ which is
  a vacuum to the operators $\{ b_h^\dagger, b_p \}$, any $N$-particle
  Slater determinant $|\Phi_1 \rangle$ which is not orthogonal to
  $|\overline{\Phi}_0 \rangle$ can be written in the form
  \begin{equation}
    |\Phi_1 \rangle =
    \mathcal{N} \, \exp \left( \sum_{ph} Z_{ph} \,
      \bar{b}_p^\dagger \, \bar{b}_h \right) |\Phi_0 \rangle,
    \label{Thou2}
  \end{equation}
  where $\mathcal{N} = \langle \overline{\Phi}_0 | \Phi_1 \rangle$ is
  a normalization constant and the coefficients $Z_{ph}$ are uniquely
  determined.
\end{quote}

For a proof of this last theorem we refer the reader to Appendix
\ref{sec:app_thou} of the present work.

\section{Matrix elements between $N$-electron Slater determinants \label{sec:matel}}

In this section we obtain the expressions required for the evaluation
of matrix elements between arbitrary Slater determinants built out of
operators resulting from a non-unitary canonical transformation.

\subsection{Norm overlaps}

The overlap between two $N$-particle Slater determinants of the form
$|\Phi_\alpha \rangle = \prod_k \alpha_k^\dagger | - \rangle$ can be
obtained by application of Wick's theorem \cite{blaizot_ripka} on the
bare fermion vacuum. That is,
\begin{equation}
  \langle \Phi_\beta | \Phi_\alpha \rangle =
  \langle -| \beta_N \cdots \beta_1 \, \alpha_1^\dagger \cdots
    \alpha_N^\dagger |-\rangle =
  \mathrm{det} \, S,
\end{equation}
where $S_{ij} = \contraction{}{\beta}{_i \,}{\alpha} \beta_i \,
\alpha_j^\dagger = \langle \beta_i | \alpha_j \rangle$. Here, we have
used the fact that the contractions $\contraction{}{\beta}{_i
  \,}{\beta} \beta_i \, \beta_j$ and
$\contraction{}{\alpha}{_i^\dagger \,}{\alpha} \alpha_i^\dagger \,
\alpha_j^\dagger$ vanish for HF-type operators.

The overlaps among $N$-particle Slater determinants become
\begin{subequations}
  \label{overlap}
  \begin{align}
    \langle \Phi_0 | \Phi_1 \rangle &=
    \mathrm{det}_N \, D^{0 \trans} \, D^{1 \ast}, \\
    \langle \Phi_0 | \overline{\Phi}_1 \rangle &=
      \mathrm{det}_N \, D^{0 \trans} \, \bar{D}^{1 \ast}, \\
    \langle \overline{\Phi}_0 | \Phi_1 \rangle &=
      \mathrm{det}_N \, \bar{D}^{0 \trans} \, D^{1 \ast}, \\
    \langle \overline{\Phi}_0 | \overline{\Phi}_1 \rangle &=
      \mathrm{det}_N \, \bar{D}^{0 \trans} \, \bar{D}^{1 \ast},
  \end{align}
\end{subequations}
where we have used $\mathrm{det}_N$ to denote that the determinant is
over the $N \times N$ set of occupied orbitals. Observe that $\langle
\Phi_0 | \overline{\Phi}_0 \rangle = \langle \overline{\Phi}_0 |
\Phi_0 \rangle = 1$, which corresponds to the bi-orthonormality
condition previously described.

\subsection{Operator matrix elements}

In deriving the expressions for operator matrix elements, we follow
Ring and Shuck \cite{ring_schuck}. Our aim in this subsection is to
evaluate matrix elements of the form
\begin{equation}
  \langle \overline{\Phi}_0 | c^\dagger_{l_1} \cdots c^\dagger_{l_p}
    \, c_{k_1} \cdots c_{k_p} | \Phi_1 \rangle. \label{matel}
\end{equation}
The form above is chosen for convenience, but other matrix elements
can be derived in the same way described below.

We shall use Thouless' theorem to write the state $|\Phi_1 \rangle$ as
\begin{align}
  |\Phi_1 \rangle &=
  \exp (\hat{\mathcal{Z}}) | \Phi_0 \rangle
    \langle \overline{\Phi}_0 | \Phi_1 \rangle, \\
  \hat{\mathcal{Z}} &=
  \sum_{ph} \mathcal{Z}_{ph} \bar{b}_p^\dagger \bar{b}_h.
\end{align}
Here, $\{ b_h^\dagger, b_p, \bar{b}_h, \bar{b}_p^\dagger \}$ are defined
such that
\begin{center}
  \begin{tabular}{c @{\hspace{1cm}} c}
    $b_h^\dagger |\Phi_0 \rangle = 0 \quad \forall \quad b_h^\dagger$, &
    $b_p |\Phi_0 \rangle = 0 \quad \forall \quad b_p$, \\
    $\langle \overline{\Phi}_0| \bar{b}_h = 0 \quad \forall \quad
      \bar{b}_h$, &
    $\langle \overline{\Phi}_0| \bar{b}_p^\dagger = 0 \quad \forall
      \quad \bar{b}_p^\dagger.$
  \end{tabular}
\end{center}

On the other hand, we write the state $\langle \overline{\Phi}_0|$ as
\begin{equation}
  \langle \overline{\Phi}_0| =
  \langle \overline{\Phi}_0| \exp (-\hat{\mathcal{Z}}),
\end{equation}
where use has been made of the vacuum properties just described.

It then follows that we can evaluate the general matrix element from
Eq. \ref{matel} as
\begin{align}
  \langle \overline{\Phi}_0 | c^\dagger_{l_1} \cdots c^\dagger_{l_p}
    \, c_{k_1} \cdots c_{k_p} | \Phi_1 \rangle
  &= \langle \overline{\Phi}_0 | \Phi_1 \rangle
  \langle \overline{\Phi}_0 | \exp (-\hat{\mathcal{Z}}) \,
    c^\dagger_{l_1} \cdots c^\dagger_{l_p} \, c_{k_1} \cdots c_{k_p}
    \, \exp (\hat{\mathcal{Z}}) |\Phi_0 \rangle \nonumber \\
  &= \langle \overline{\Phi}_0 | \Phi_1 \rangle
    \langle \overline{\Phi}_0 | \tilde{d}_{l_1} \cdots
    \tilde{d}_{l_p} \, d_{k_1} \cdots d_{k_p} |\Phi_0 \rangle,
\end{align}
where we have introduced the operators
\begin{subequations}
  \begin{align}
    \tilde{d}_l &=
    \exp (-\hat{\mathcal{Z}}) \, c_l^\dagger \,
      \exp (\hat{\mathcal{Z}}), \\
    d_k &=
    \exp (-\hat{\mathcal{Z}}) \, c_k \,
      \exp (\hat{\mathcal{Z}}).
  \end{align}      
\end{subequations}

We now express the operators $\{ \tilde{d}_l, d_k \}$ in terms of $\{
b_h^\dagger, b_p, \bar{b}_h, \bar{b}_p^\dagger \}$. This is
accomplished by using Eq. \ref{HFinv} to write $\{ c_j, c_j^\dagger
\}$ in terms of $\{ b_h^\dagger, b_p, \bar{b}_h, \bar{b}_p^\dagger
\}$. It follows that
\begin{align}
  \tilde{d}_l
  &= \exp (-\hat{\mathcal{Z}}) \, c_l^\dagger \,
    \exp (\hat{\mathcal{Z}})
  = c_l^\dagger - \left[ \hat{\mathcal{Z}}, c_l^\dagger \right]
    \nonumber \\
  &= \sum_h \bar{D}^0_{lh} \, b_h^\dagger + \sum_p \left( D^0_{lp} -
    \sum_h \mathcal{Z}_{ph} \, \bar{D}^0_{lh}\right) \,
    \bar{b}_p^\dagger,
\end{align}
\begin{align}
  d_k
  &= \exp (-\hat{\mathcal{Z}}) \, c_k \,
    \exp (\hat{\mathcal{Z}})
  = c_k - \left[ \hat{\mathcal{Z}}, c_k \right]
    \nonumber \\
  &= \sum_h \left( D_{kh}^{0 \ast} + \sum_p \mathcal{Z}_{ph} \,
    \bar{D}_{kp}^{0 \ast} \right) \, \bar{b}_h + \sum_p
    \bar{D}_{kp}^{0 \ast} \, b_p.
\end{align}

Because $\{ \tilde{d}_l, d_k \}$ are given as linear combinations of
$\{ b_h^\dagger, b_p, \bar{b}_h, \bar{b}_p^\dagger \}$, Wick's theorem
\cite{blaizot_ripka} can be used to calculate the corresponding matrix
elements.  The non-vanishing contractions among the operators $\{
b_h^\dagger, b_p, \bar{b}_h, \bar{b}_p^\dagger \}$ are given by
\begin{subequations}
  \begin{align}
    \contraction{}{b}{_h^\dagger \,}{\bar{b}} b_h^\dagger \,
      \bar{b}_{h'} &= \delta_{hh'}, \\
    \contraction{}{b}{_p \,}{\bar{b}} b_p \,
      \bar{b}_{p'}^\dagger &= \delta_{pp'}.
  \end{align}
\end{subequations}

It follows that the non-vanishing contractions among the operators $\{
\tilde{d}_l, d_k \}$ are of the form
\begin{align}
  \contraction{}{\tilde{d}}{_l \,}{d} \tilde{d}_l \, d_k
  &= \sum_{hh'} \bar{D}^0_{lh} \, \left( D^{0 \ast}_{kh'} + \sum_p
    \mathcal{Z}_{ph'} \, \bar{D}^{0 \ast}_{kp} \right) \, \delta_{hh'}
    \nonumber \\
  &= \sum_h \bar{D}^0_{lh} \, D_{kh}^{0 \ast} + \sum_{ph}
    \bar{D}^0_{lh} \, \mathcal{Z}_{ph} \, \bar{D}_{kp}^{0 \ast},
\end{align}
\begin{align}
  \contraction{}{d}{_l \,}{\tilde{d}} d_l \, \tilde{d}_k
  &= \sum_{pp'} \bar{D}_{lp}^{0 \ast} \, \left( D^0_{kp'} - \sum_h
    \mathcal{Z}_{p'h} \, \bar{D}^0_{kh} \right) \, \delta_{pp'}
    \nonumber \\
  &= \sum_p \bar{D}_{lp}^{0 \ast} \, D^0_{kp} - \sum_{ph}
    \bar{D}_{lp}^{0 \ast} \, \mathcal{Z}_{ph} \, \bar{D}^0_{kh}.
\end{align}

The application of Wick's theorem to the operator matrix elements of
the form of Eq. \ref{matel} leads us to conclude that all such matrix
elements can be evaluated in terms of the transition density matrix
$\rho^{\bar{0}1}$, given by
\begin{align}
  \rho^{\bar{0}1}_{kl}
  &= \frac{\langle \overline{\Phi}_0 | c_l^\dagger \, c_k | \Phi_1 \rangle}%
          {\langle \overline{\Phi}_0 | \Phi_1 \rangle}
  = \langle \overline{\Phi}_0| \exp(-\hat{\mathcal{Z}}) \, c_l^\dagger
    \, c_k \, \exp(\hat{\mathcal{Z}}) |\Phi_0 \rangle \nonumber \\
  &= \sum_h \bar{D}^0_{lh} \, D_{kh}^{0 \ast} + \sum_{ph}
    \bar{D}^0_{lh} \, \mathcal{Z}_{ph} \, \bar{D}_{kp}^{0 \ast},
\end{align}
where
\begin{align}
  \mathcal{Z}_{ph} &=
  \sum_{h'} \left( D^{0 \trans} \, D^{1 \ast} \right)_{ph'} \left(
    \mathcal{L}^{\ast -1} \right)_{h'h}, \\
  \mathcal{L}_{h'h} &=
  \left( \bar{D}^{0 \dagger} \, D^1 \right)_{h'h}.
\end{align}
Here, we have used Eqs. \ref{defLMY} and \ref{defZ} from Appendix
\ref{sec:app_thou} to write the forms of the matrices $\mathcal{Z}$
and $\mathcal{L}$.

\subsection{Evaluation of the energy of a single Slater determinant}

As an example of the application of the above equations, let us now
consider the evaluation of the energy of a determinant $|\Phi
\rangle$. Given a two-body Hamiltonian in the usual second-quantized
form \cite{blaizot_ripka}
\begin{equation}
  \hat{H} = \sum_{ik} \langle i | \hat{h} | k \rangle \, c_i^\dagger
    \, c_k + \frac{1}{4} \sum_{ijkl} \langle ij | \hat{v} | kl \rangle
    \, c_i^\dagger \, c_j^\dagger \, c_l \, c_k,
\end{equation}
where $\langle i | \hat{h} | k \rangle$ and $\langle ij | \hat{v} | kl
\rangle$ are one- and anti-symmetrized two-particle integrals,
respectively, the energy can be evaluated as
\begin{align}
  E &= \frac{\langle \Phi | \hat{H} | \Phi \rangle}{\langle \Phi
    | \Phi \rangle} \nonumber \\
  &= \sum_{ik} h_{ik} \, \rho_{ki} + \frac{1}{2} \sum_{ijkl} \langle
    ij | \hat{v} | kl \rangle \, \rho_{ki} \, \rho_{lj} \nonumber \\
  &= \mathrm{Tr} \left( h \, \rho + \frac{1}{2} \, \Gamma \, \rho
    \right), \label{enerSD}
\end{align}
where
\begin{align}
  \rho_{ki}
  &= \frac{\langle \Phi | c_i^\dagger \, c_k | \Phi \rangle}{\langle
    \Phi | \Phi \rangle} \nonumber \label{HFrho} \\
  &= \sum_h D_{ih} \, \bar{D}_{kh}^\ast + \sum_{ph} D_{ih} \,
    \bar{\mathcal{Z}}_{ph} \, D_{kp}^\ast, \\
  \Gamma_{ik} &= \sum_{jl} \langle ij | \hat{v} | kl \rangle \,
    \rho_{lj},
\end{align}
and
\begin{align}
  \bar{\mathcal{Z}}_{ph} &= \sum_{h'} \left( \bar{D}^\trans \, D^\ast
    \right)_{ph'} \left( \bar{\mathcal{L}}^{\ast -1} \right)_{h'h}, \\
  \bar{\mathcal{L}}_{h'h} &= \left( D^\dagger \, D \right)_{h'h}.
\end{align}

It is important to realize that the energy expression
(Eq. \ref{enerSD}) has the same form as in standard ({\em i.e.}
unitary) HF. The difference lies in the form of the density matrix
$\rho$ (Eq. \ref{HFrho}), which comes about from the fact that the
anti-commutation relations satisfied by the HF operators are
different.

\section{Variational ansatz with Slater determinants from non-unitary transformations \label{sec:ansatz}}

In this section, we use a simple, two-determinant ansatz that uses the full
flexibility of the non-unitary HF-like transformation of
Eq. \ref{HFtrans} as part of a variational strategy.

Before introducing such ansatz, we note that using a single Slater
determinant $|\Phi \rangle$ as a trial wavefunction, whether resulting
from a unitary or a non-unitary canonical transformation, would lead
to the same variational energy. An $N$-particle Slater determinant
resulting from a non-unitary canonical transformation is equivalent to
an un-normalized Slater determinant in the usual ({\em i.e.} unitary)
sense. The variational optimization of the energy (taken as the
Hamiltonian overlap over the norm overlap) would lead to the same
result regardless of the underlying normalization of the determinant.

The two-determinant ansatz that we use is given by
\begin{align}
  |\Psi \rangle &= c_1 |\Phi \rangle + c_2 |\overline{\Phi} \rangle,
    \nonumber \\
  &\equiv c_1 |\Phi_1 \rangle + c_2 |\Phi_2 \rangle,
  \label{defPsi}
\end{align}
where $c_1$ and $c_2$ are coefficients to be determined
variationally. We have made the identification $|\Phi_1 \rangle \equiv
|\Phi \rangle$ and $|\Phi_2 \rangle \equiv |\overline{\Phi} \rangle$
to simplify our notation below. Observe that for a standard ({\em
  i.e.}  unitary) HF transformation, $|\Phi_1 \rangle =
|\overline{\Phi} \rangle$, which in turn implies $|\Psi \rangle =
|\Phi \rangle$.

One could argue that the ansatz of Eq. \ref{defPsi} has the same
variational flexibility as that in which $|\Phi_1\rangle$ and $|\Phi_2
\rangle$ are two non-orthogonal Slater determinants resulting, each of
them, from a standard unitary canonical transformation (see
Ref. \cite{tomita2004}). Nevertheless, the ansatz we use explicitly
results from a single linear canonical transformation of the
elementary fermion operators.

The Hamiltonian expectation value associated with the state $|\Psi
\rangle$ is given by
\begin{equation}
  E =
  \frac{\displaystyle \sum_{\alpha, \beta=1}^2 c_\alpha^\ast \,
         c_\beta \langle \Phi_\alpha | \hat{H} | \Phi_\beta \rangle}%
       {\displaystyle \sum_{\alpha,\beta=1}^2 c_\alpha^\ast \,
         c_\beta \langle \Phi_\alpha | \Phi_\beta \rangle}.
\end{equation}

We rewrite the energy above in the form
\begin{align}
  E &=
  \sum_{\alpha,\beta=1}^2 y_{\alpha \beta}
  \frac{\langle \Phi_\alpha | \hat{H} | \Phi_\beta \rangle}%
       {\langle \Phi_\alpha | \Phi_\beta \rangle},
    \label{EPsi} \\[4pt]
  y_{\alpha \beta} &=
  \frac{c_\alpha^\ast \, c_\beta \langle \Phi_\alpha | \Phi_\beta \rangle}%
       {\displaystyle \sum_{\alpha',\beta'=1}^2 c_{\alpha'}^\ast \,
         c_{\beta'} \langle \Phi_{\alpha'} | \Phi_{\beta'} \rangle}.
    \label{yPsi}
\end{align}

The matrix elements appearing in Eqs. \ref{EPsi} and \ref{yPsi} can be
evaluated in a straight-forward way. The overlap kernels in
Eq. \ref{yPsi} are computed as
\begin{subequations}
  \begin{align}
    \langle \Phi | \Phi \rangle &=
      \mathrm{det}_N \, D^\trans \, D^\ast, \\
    \langle \overline{\Phi} | \Phi \rangle &=
      \mathrm{det}_N \, \bar{D}^\trans \, D^\ast = 1, \\
    \langle \Phi | \overline{\Phi} \rangle &=
      \mathrm{det}_N \, D^\trans \, \bar{D}^\ast = 1, \\
    \langle \overline{\Phi} | \overline{\Phi} \rangle &=
      \mathrm{det}_N \, \bar{D}^\trans \, \bar{D}^\ast.
  \end{align}
\end{subequations}

The Hamiltonian kernels are evaluated in terms of transition density
matrices as
\begin{align}
  \frac{\langle \Phi_\alpha | \hat{H} | \Phi_\beta \rangle}%
       {\langle \Phi_\alpha | \Phi_\beta \rangle} &=
  \mathrm{Tr} \left( h \, \rho^{\alpha \beta} +
    \frac{1}{2} \, \Gamma^{\alpha \beta} \, \rho^{\alpha \beta} \right), \\
  \Gamma_{ik}^{\alpha \beta} &=
  \sum_{jl} \langle ij | \hat{v} | kl \rangle \, \rho_{lj}^{\alpha \beta}.
\end{align}

The transition density matrices are in turn given by
\begin{subequations}
  \begin{align}
    \rho_{ki}^{11} &=
    \frac{\langle \Phi | c_i^\dagger \, c_k | \Phi \rangle}%
         {\langle \Phi | \Phi \rangle}
    = \sum_h D_{ih} \, \bar{D}_{kh}^\ast + \sum_{ph} D_{ih} \,
      \bar{\mathcal{Z}}_{ph} \, D_{kp}^\ast, \\[4pt]
    \rho_{ki}^{21} &=
    \frac{\langle \overline{\Phi} | c_i^\dagger \, c_k | \Phi \rangle}%
         {\langle \overline{\Phi} | \Phi \rangle}
    = \sum_h \bar{D}_{ih} \, D_{kh}^\ast, \\[4pt]
    \rho_{ki}^{12} &=
    \frac{\langle \Phi | c_i^\dagger \, c_k | \overline{\Phi} \rangle}%
         {\langle \Phi | \overline{\Phi} \rangle}
    = \sum_h D_{ih} \, \bar{D}_{kh}^\ast, \\[4pt]
    \rho_{ki}^{22} &=
   \frac{\langle \overline{\Phi} | c_i^\dagger \, c_k | \overline{\Phi} \rangle}%
        {\langle \overline{\Phi} | \overline{\Phi} \rangle}
    = \sum_h \bar{D}_{ih} \, D_{kh}^\ast + \sum_{ph} \bar{D}_{ih} \,
      \mathcal{Z}_{ph} \, \bar{D}_{kp}^\ast.
  \end{align}
\end{subequations}
Here,
\begin{subequations}
  \begin{align}
    \bar{\mathcal{Z}}_{ph} &=
    \sum_{h'} \left( \bar{D}^\trans \, D^\ast \right)_{ph'}
      \left( \bar{\mathcal{L}}^{\ast -1} \right)_{h'h}, \\
    \mathcal{Z}_{ph} &=
    \sum_{h'} \left( D^\trans \, \bar{D}^\ast \right)_{ph'}
      \left( \mathcal{L}^{\ast -1} \right)_{h'h}, \\
    \bar{\mathcal{L}}_{h'h} &=
    \left( D^\dagger \, D \right)_{h'h}, \\
    \mathcal{L}_{h'h} &=
    \left( \bar{D}^\dagger \, \bar{D} \right)_{h'h}.
  \end{align}
\end{subequations}

\subsection{Variational optimization of $|\Phi \rangle$ \label{sec:opt}}

Let us now consider the variational optimization of the wavefunction
ansatz introduced in Eq. \ref{defPsi}. The variational parameters are
the coefficients $c_1$ and $c_2$ and the orbital coefficients (that
is, the matrices $D$ and $\bar{D}$) defining the states $|\Phi \rangle$
and $|\overline{\Phi} \rangle$. The variation has to be carried out
subject to the constraint that $\langle \overline{\Phi} | \Phi \rangle
= 1$, which is equivalent to saying that $|\Phi \rangle$ and
$|\overline{\Phi} \rangle$ are defined by a canonical transformation
of the form of Eq. \ref{HFtrans}.

The variation with respect to the coefficients $c_1$ and $c_2$ yields
the generalized eigenvalue problem
\begin{equation}
  ( \mathbf{H} - E \, \mathbf{N} ) \, \boldsymbol{c} = 0, \label{gev1}
\end{equation}
with the constraint
\begin{equation}
  \boldsymbol{c}^\dagger \, \mathbf{N} \, \boldsymbol{c} = 1, \label{gev2}
\end{equation}
which ensures the orthonormality of the solution. Here,
$\boldsymbol{c}$ represents the column of coefficients $\{ c_1, c_2
\}$, while $\mathbf{H}$ and $\mathbf{N}$ are, respectively,
Hamiltonian and overlap matrices given by
\begin{align}
  H_{\alpha \beta} &=
  \langle \Phi_\alpha | \hat{H} | \Phi_\beta \rangle, \\
  N_{\alpha \beta} &=
  \langle \Phi_\alpha | \Phi_\beta \rangle.
\end{align}
It should be stressed that at this level we only keep the
lowest-energy solution to the generalized eigenvalue problem, in a
similar way as in projected-HF methods involving an eigenvalue problem
\cite{rodriguez-guzman2012}.

Let us now consider the variation in the energy with respect to the
underlying non-unitary HF transformation. We have followed the work of
Egido and coworkers \cite{egido1995} for this purpose. Let us assume
that we are provided a guess for $|\Phi \rangle$ and $|\overline{\Phi}
\rangle$, characterized by the set of HF operators $\{ b_h^\dagger,
b_p, \bar{b}_h, \bar{b}_p^\dagger \}$. We can now parametrize the
energy functional around $\{ |\Phi \rangle, |\overline{\Phi} \rangle
\}$ by allowing for independent Thouless' rotations of both states,
characterized by the matrices $Z$ and $\bar{Z}$. That is, we let
\begin{subequations}
  \begin{align}
    |\Phi \rangle &\rightarrow \exp \left( \sum_{ph} Z_{ph} \,
      \bar{b}_p^\dagger \, \bar{b}_h \right) |\Phi \rangle,
      \label{rotphi} \\[4pt]
    |\overline{\Phi} \rangle &\rightarrow \exp \left( \sum_{ph}
      \bar{Z}_{ph} \, b_p^\dagger \, b_h \right) |\overline{\Phi}
      \rangle.
      \label{rotphib}
  \end{align}
\end{subequations}

We define the local gradient $\{ G, \bar{G} \}$ around $Z = 0$ and
$\bar{Z} = 0$ as
\begin{subequations}
  \begin{align}
    G_{ph} &= - \left. \frac{\partial}{\partial \, Z_{ph}^\ast} \, E
      \, [Z, \bar{Z}] \right|_{Z_{ph} = 0}, \label{LocalG1} \\
    \bar{G}_{ph} &= - \left. \frac{\partial}{\partial \,
      \bar{Z}_{ph}^\ast} \, E \, [Z, \bar{Z}] \right|_{\bar{Z}_{ph} =
      0}. \label{LocalG2}
  \end{align}
\end{subequations}
Here, $Z_{ph}$ and $Z^\ast_{ph}$ are treated as independent
variables, and the same is true for $\bar{Z}_{ph}$ and
$\bar{Z}^\ast_{ph}$. The total derivative of the energy then becomes
\begin{equation}
  d E = - \sum_{ph} \left[ G_{ph} \, d Z_{ph}^\ast + \bar{G}_{ph} \,
    d \bar{Z}_{ph}^\ast + \mathrm{c.c.}  \right].
\end{equation}

Explicit differentiation of the parametrized energy functional leads
to the following expressions for the local gradient:
\begin{subequations}
  \label{grad}
  \begin{align}
    G_{ph} &=
    - y_{11} \, \frac{\langle \Phi | \bar{b}_h^\dagger \, \bar{b}_p \,
      \left( \hat{H} - E \right) | \Phi \rangle}{\langle \Phi | \Phi
      \rangle}
    - y_{12} \, \frac{\langle \Phi | \bar{b}_h^\dagger \, \bar{b}_p \,
      \left( \hat{H} - E \right) | \overline{\Phi} \rangle}{\langle \Phi
      | \overline{\Phi} \rangle}, \\[4pt]
    \bar{G}_{ph} &=
    - y_{21} \, \frac{\langle \overline{\Phi} | b_h^\dagger \, b_p \,
      \left( \hat{H} - E \right) | \Phi \rangle}{\langle \overline{\Phi}
      | \Phi \rangle}
    - y_{22} \, \frac{\langle \overline{\Phi} | b_h^\dagger \, b_p \,
      \left( \hat{H} - E \right) | \overline{\Phi} \rangle}%
      {\langle \overline{\Phi} | \overline{\Phi} \rangle},
  \end{align}
\end{subequations}
where $E$ is the energy corresponding to the state $|\Psi \rangle$
from Eq. \ref{defPsi}.

The overlap-like matrix elements appearing in Eq. \ref{grad} can be
evaluated as
\begin{subequations}
  \begin{align}
    \frac{\langle \Phi | \bar{b}_h^\dagger \, \bar{b}_p | \Phi \rangle}%
         {\langle \Phi | \Phi \rangle} &=
    \sum_{mn} \bar{D}_{mh}^\ast \, \bar{D}_{np} \, \rho^{11}_{nm},
    \\[4pt]
    \frac{\langle \overline{\Phi} | b_h^\dagger \, b_p | \Phi \rangle}%
         {\langle \overline{\Phi} | \Phi \rangle} &=
    0,
    \\[4pt]
    \frac{\langle \Phi | \bar{b}_h^\dagger \, \bar{b}_p | \overline{\Phi} \rangle}%
         {\langle \Phi | \overline{\Phi} \rangle} &=
    0,
    \\[4pt]
    \frac{\langle \overline{\Phi} | b_h^\dagger \, b_p | \overline{\Phi} \rangle}%
         {\langle \overline{\Phi} | \overline{\Phi} \rangle} &=
    \sum_{mn} D_{mh}^\ast \, D_{np} \, \rho^{22}_{nm}.
  \end{align}
\end{subequations}

The Hamiltonian-like matrix elements appearing in Eq. \ref{grad} can
be evaluated as
\begin{subequations}
  \begin{align}
    \frac{\langle \Phi | \bar{b}_h^\dagger \, \bar{b}_p \, \hat{H} | \Phi \rangle}%
         {\langle \Phi | \Phi \rangle} &=
    \sum_{mn} \bar{D}_{mh}^\ast \, \bar{D}_{np} \, \rho^{11}_{nm} \,
    \frac{\langle \Phi | \hat{H} | \Phi \rangle}%
         {\langle \Phi | \Phi \rangle} \nonumber \\ &+
    \sum_{mn} \sum_{ik} \bar{D}_{mh}^\ast \, \bar{D}_{np} \, \left(
      h_{ik} + \Gamma^{11}_{ik} \right) \, \rho^{11}_{km} \, \left(
      \delta_{ni} - \rho^{11}_{ni} \right),
    \\[4pt]
    \frac{\langle \overline{\Phi} | b_h^\dagger \, b_p \, \hat{H} | \Phi \rangle}%
         {\langle \overline{\Phi} | \Phi \rangle} &=
    \sum_{ik} D_{kh}^\ast \, D_{ip} \, \left( h_{ik} + \Gamma^{21}_{ik}
      \right),
    \\[4pt]
    \frac{\langle \Phi | \bar{b}_h^\dagger \, \bar{b}_p \, \hat{H} | \overline{\Phi} \rangle}%
         {\langle \Phi | \overline{\Phi} \rangle} &=
    \sum_{ik} \bar{D}_{kh}^\ast \, \bar{D}_{ip} \, \left( h_{ik} +
      \Gamma^{12}_{ik} \right),
    \\[4pt]
    \frac{\langle \overline{\Phi} | b_h^\dagger \, b_p \, \hat{H} | \overline{\Phi} \rangle}%
         {\langle \overline{\Phi} | \overline{\Phi} \rangle} &=
    \sum_{mn} D_{mh}^\ast \, D_{np} \, \rho^{22}_{nm} \,
    \frac{\langle \overline{\Phi} | \hat{H} | \overline{\Phi} \rangle}%
         {\langle \overline{\Phi} | \overline{\Phi} \rangle} \nonumber \\ &+
    \sum_{mn} \sum_{ik} D_{mh}^\ast \, D_{np} \, \left( h_{ik} +
      \Gamma^{22}_{ik} \right) \, \rho^{22}_{km} \, \left( \delta_{ni} -
      \rho^{22}_{ni} \right).
  \end{align}
\end{subequations}

\subsection{Restoration of the bi-orthonormality condition}

Let us assume that, during the optimization process, we started with
the states $|\Phi \rangle$ and $|\overline{\Phi} \rangle$ and produced
the new states $|\Phi' \rangle$ and $|\overline{\Phi'} \rangle$ by
using the Thouless' transformations
\begin{subequations}
  \begin{align}
    |\Phi' \rangle &=
    \mathcal{N} \, \exp \left( \sum_{ph} Z_{ph} \, \bar{b}_p^\dagger
      \, \bar{b}_h \right) |\Phi \rangle, \\[4pt]
    |\overline{\Phi'} \rangle &=
    \mathcal{\bar{N}} \, \exp \left( \sum_{ph} \bar{Z}_{ph} \,
      b_p^\dagger \, b_h \right) |\overline{\Phi} \rangle.
  \end{align}
\end{subequations}
Here, the matrices $Z$ and $\bar{Z}$ can be chosen as, for instance,
\begin{subequations}
  \begin{align}
    Z_{ph} &= \eta \, G_{ph}, \\
    \bar{Z}_{ph} &= \eta \, \bar{G}_{ph},
  \end{align}
\end{subequations}
with $\eta \geq 0$ being some parameter. We denote with $\{
\tilde{d}_h^\dagger, \tilde{d}_p, \tilde{\bar{d}}_h,
\tilde{\bar{d}}_p^\dagger \}$ the set of HF operators produced by such
transformations (see Eqs. \ref{th1} and \ref{th2})
\begin{subequations}
  \begin{align}
    \tilde{d}_h^\dagger &= b_h^\dagger + \sum_{p} Z_{ph} \,
      \bar{b}_p^\dagger, \\
    \tilde{d}_p &= b_p - \sum_{h} Z_{ph} \, \bar{b}_h, \\
    \tilde{\bar{d}}_h &= \bar{b}_h + \sum_{p} \bar{Z}_{ph}^\ast \,
      b_p, \\
    \tilde{\bar{d}}_p^\dagger &= \bar{b}_p^\dagger - \sum_{h}
      \bar{Z}_{ph}^\ast \, b_h^\dagger,
  \end{align}
\end{subequations}
where the operators $\{ b_h^\dagger, b_p, \bar{b}_h, \bar{b}_p^\dagger
\}$ describing the states $|\Phi \rangle$ and $|\overline{\Phi}
\rangle$ are assumed to satisfy all the appropriate anti-commutation
relations.

We show in Appendix \ref{sec:app_thou} that the operators $\{
\tilde{d}_h^\dagger, \tilde{d}_p \}$ annihilate the vacuum $|\Phi'
\rangle$. Similarly, the operators $\{ \tilde{\bar{d}}_h^\dagger,
\tilde{\bar{d}}_p \}$ annihilate the vacuum $|\overline{\Phi'}
\rangle$. The operators $\{ \tilde{d}_h^\dagger, \tilde{d}_p,
\tilde{\bar{d}}_h, \tilde{\bar{d}}_p^\dagger \}$ do not, however,
satisfy the anti-commutation relations given by Eq. \ref{HFanticom}.
In fact, they satisfy
\begin{subequations}
  \begin{align}
    \Big[ \tilde{d}_p, \tilde{d}_h^\dagger \Big]_+ &= 0, \\
    \Big[ \tilde{\bar{d}}_h, \tilde{\bar{d}}_p^\dagger \Big]_+ &= 0 \\
    \Big[ \tilde{\bar{d}}_h, \tilde{d}_{h'}^\dagger \Big]_+ &=
      \left( I +  Z^\trans \, \bar{Z}^\ast \right)_{h'h}, \\
    \Big[ \tilde{d}_p, \tilde{\bar{d}}_{p'}^\dagger \Big]_+ &=
      \left( I + \bar{Z}^\ast \, Z^\trans \right)_{p'p}.
  \end{align}
\end{subequations}

We can restore the desired anti-commutation relations by performing
the transformations
\begin{subequations}
  \begin{align}
    d_h^\dagger &= \sum_{h'} L_{hh'}^{-1} \, \tilde{d}_{h'}^\dagger, \\
    \bar{d}_h &= \sum_{h'} \bar{L}_{hh'}^{\ast -1} \,
      \tilde{\bar{d}}_{h'}, \\
    d_p &= \sum_{p'} M_{pp'}^{\ast -1} \, \tilde{d}_{p'}, \\
    \bar{d}_p^\dagger &= \sum_{p'} \bar{M}_{pp'}^{-1} \,
      \tilde{\bar{d}}_{p'}^\dagger,
  \end{align}
\end{subequations}
in terms of the lower triangular matrices $L$, $\bar{L}$, $M$, and
$\bar{M}$ \cite{egido1995}.

The anti-commutation relations among $\{ d_h^\dagger, d_p, \bar{d}_h,
\bar{d}_p^\dagger \}$ become
\begin{subequations}
  \begin{align}
    \Big[ \bar{d}_h, d_{h'}^\dagger \Big]_+ &=
      \sum_{\mu \nu} \bar{L}_{h \mu}^{\ast -1} \, L_{h' \nu}^{-1} \,
      \left( I +  Z^\trans \, \bar{Z}^\ast \right)_{\nu \mu} =
      \delta_{h'h}, \\
    \Big[ \tilde{d}_p, \tilde{\bar{d}}_{p'}^\dagger \Big]_+ &=
      \sum_{\mu \nu} M_{p \mu}^{\ast -1} \, \bar{M}_{p' \nu}^{-1} \,
      \left( I + \bar{Z}^\ast \, Z^\trans \right)_{\nu \mu} =
      \delta_{p'p},
  \end{align}
\end{subequations}
which yield the following equations for determining $L$, $\bar{L}$,
$M$, and $\bar{M}$:
\begin{subequations}
  \begin{align}
    I + Z^\trans \, \bar{Z}^\ast &= L \, \bar{L}^\dagger,
      \label{lu1} \\
    I + \bar{Z}^\ast \, Z^\trans &= \bar{M} \, M^\dagger.
      \label{lu2}
  \end{align}
\end{subequations}
Hence, given the matrices $Z$ and $\bar{Z}$, standard LU
decompositions (Eqs. \ref{lu1} and \ref{lu2}) can be performed to
obtain the matrices $L$, $\bar{L}$, $M$, and $\bar{M}$. This is
similar to the unitary case, where the only two matrices required ($L$
and $M$) can be obtained by Cholesky decompositions
\cite{schmid2005,rodriguez-guzman2012}.

We remark that if $\bar{Z} = 0$ (or $Z = 0$), then the operators $\{
\tilde{d}_h^\dagger, \tilde{d}_p, \tilde{\bar{d}}_h,
\tilde{\bar{d}}_p^\dagger \}$ do obey all the required
anti-commutation relations. In other words, one has to restore the
bi-orthonormality condition only if both $|\Phi \rangle$ and
$|\overline{\Phi} \rangle$ are rotated.

\subsection{Global gradient}

In order to use gradient-based optimization methods such as the
conjugate gradient or quasi-Newton methods (see
Refs. \onlinecite{rodriguez-guzman2012},
\onlinecite{rodriguez-guzman2004}, \onlinecite{rodriguez-guzman2004b},
\onlinecite{egido1995}, and \onlinecite{schmid2005}), one must be able
to compute a global gradient. That is, we should be able to compute
the gradient of the energy at $|\Psi_1 \rangle$ with respect to
variations in $Z$ and $\bar{Z}$ defined in terms of the operators $\{
b_h^{0 \dagger}, b^0_p, \bar{b}^0_h, \bar{b}_p^{0 \dagger} \}$
corresponding to the reference state $|\Psi_0 \rangle$. Here, we
follow Egido {\em et al}.  \cite{egido1995} in deriving the form of
the global gradient.

Consider the energy of the state $|\Psi_1 \rangle$. It is given by
\begin{equation}
  E [\Psi_1] =
  \frac{\displaystyle \sum_{\alpha, \beta=1}^2 c_\alpha^\ast \,
         c_\beta \langle \Phi^1_\alpha | \hat{H} | \Phi^1_\beta \rangle}%
       {\displaystyle \sum_{\alpha,\beta=1}^2 c_\alpha^\ast \,
         c_\beta \langle \Phi^1_\alpha | \Phi^1_\beta \rangle}.
\end{equation}

Provided that $|\Phi^1 \rangle$ and $|\overline{\Phi}^1 \rangle$ are
non-orthogonal to $\langle \overline{\Phi}^0|$ and $\langle \Phi^0|$,
respectively, we can write
\begin{subequations}
  \begin{align}
    |\Phi^1 \rangle &=
    \mathcal{N} \, \exp \left( \sum_{ph} Z_{ph} \, \bar{b}_p^{0
      \dagger} \, \bar{b}^0_h \right) |\Phi^0 \rangle, \\[4pt]
    |\overline{\Phi}^1 \rangle &=
    \bar{\mathcal{N}} \, \exp \left( \sum_{ph} \bar{Z}_{ph} \, b_p^{0
      \dagger} \, b^0_h \right) |\overline{\Phi}^0 \rangle,
  \end{align}
\end{subequations}
where $\mathcal{N} = \langle \overline{\Phi}^0 | \Phi^1 \rangle$ and
$\bar{\mathcal{N}} = \langle \Phi^0 | \overline{\Phi}^1 \rangle$ are
normalization constants. Here,
\begin{subequations}
  \begin{align}
    Z_{ph} &= \sum_{h'} \left( D^{0 \trans} \, D^{1 \ast}
      \right)_{ph'} \left( \mathcal{L}^{\ast -1} \right)_{h'h}, \\
    \bar{Z}_{ph} &= \sum_{h'} \left( \bar{D}^{0 \trans} \, \bar{D}^{1
      \ast} \right)_{ph'} \left( \bar{\mathcal{L}}^{\ast -1}
      \right)_{h'h}, \\
    \mathcal{L}_{h'h} &= \left( \bar{D}^{0 \dagger} \, D^1
      \right)_{h'h}, \\
    \bar{\mathcal{L}}_{h'h} &= \left( D^{0 \dagger} \, \bar{D}^1
      \right)_{h'h},
  \end{align}
\end{subequations}
where we have used Eqs. \ref{defLMY} and \ref{defZ} to write $Z$ and
$\bar{Z}$ in terms of the matrices of orbital coefficients $D^0$,
$\bar{D}^0$, $D^1$, and $\bar{D}^1$.

A variation in $Z$ and $\bar{Z}$ leads to a change in energy given by
\begin{align}
  \delta E
  &= \sum_{ph} \left[ \frac{\partial E}{\partial Z_{ph}^\ast} \, \delta
      Z_{ph}^\ast + \frac{\partial E}{\partial \bar{Z}_{ph}^\ast} \,
      \delta \bar{Z}_{ph}^\ast \right] + \mathrm{c.c.} \nonumber \\
  &= \sum_{ph} \left[ -\mathcal{G}_{ph} \, \delta
      Z_{ph}^\ast - \bar{\mathcal{G}}_{ph} \,
      \delta \bar{Z}_{ph}^\ast \right] + \mathrm{c.c.},
\end{align}
where we have introduced the global gradients $\mathcal{G}$ and
$\mathcal{\bar{G}}$ given by
\begin{subequations}
  \label{ggrad}
  \begin{align}
    \mathcal{G}_{ph} &=
    - y_{11} \, \frac{\langle \Phi^1 | \bar{b}_h^{0 \dagger} \,
      \bar{b}^0_p \, \left( \hat{H} - E \right) | \Phi^1
      \rangle}{\langle \Phi^1 | \Phi^1 \rangle}
    - y_{12} \, \frac{\langle \Phi^1 | \bar{b}_h^{0 \dagger} \,
      \bar{b}^0_p \, \left( \hat{H} - E \right) | \overline{\Phi}^1
      \rangle}{\langle \Phi^1 | \overline{\Phi}^1 \rangle}, \\[4pt]
    \bar{\mathcal{G}}_{ph} &=
    - y_{21} \, \frac{\langle \overline{\Phi}^1 | b_h^{0 \dagger} \,
      b^0_p \, \left( \hat{H} - E \right) | \Phi^1 \rangle}{\langle
      \overline{\Phi}^1 | \Phi^1 \rangle}
    - y_{22} \, \frac{\langle \overline{\Phi}^1 | b_h^{0 \dagger} \,
      b^0_p \, \left( \hat{H} - E \right) | \overline{\Phi}^1
      \rangle}{\langle \overline{\Phi}^1 | \overline{\Phi}^1
      \rangle}.
  \end{align}
\end{subequations}

In order to evaluate the matrix elements appearing in the global
gradient (Eq. \ref{ggrad}), we need to relate the operators $\{ b_h^{0
  \dagger}, b^0_p, \bar{b}^0_h, \bar{b}_p^{0 \dagger} \}$ to the
operators $\{ b_h^{1 \dagger}, b^1_p, \bar{b}^1_h, \bar{b}_p^{1
  \dagger} \}$. Combining the results of the previous subsection with
Eqs. \ref{th1} and \ref{th2}, we arrive at
\begin{subequations}
  \begin{align}
    b_h^{1 \dagger}
    &= \sum_{h'} L_{hh'}^{-1} \, \tilde{b}_{h'}^{0 \dagger}
    = \sum_{h'} L_{hh'}^{-1} \, \left( b_{h'}^{0 \dagger} + \sum_p
      Z_{ph'} \, \bar{b}_p^{0 \dagger} \right), \label{ggop1} \\
    b_p^1
    &= \sum_{p'} M_{pp'}^{\ast -1} \, \tilde{b}_{p'}^0
    = \sum_{p'} M_{pp'}^{\ast -1} \, \left( b_{p'}^0 - \sum_h
      Z_{p'h} \, \bar{b}_h^0 \right), \label{ggop2} \\
    \bar{b}_h^1
    &= \sum_{h'} \bar{L}_{hh'}^{\ast -1} \, \tilde{\bar{b}}_{h'}^0
    = \sum_{h'} \bar{L}_{hh'}^{\ast -1} \, \left( \bar{b}_{h'}^0 +
      \sum_{p} \bar{Z}_{ph'}^\ast \, b_p^0 \right), \label{ggop3} \\
    \bar{b}_p^{1 \dagger}
    &= \sum_{p'} \bar{M}_{pp'}^{-1} \, \tilde{\bar{b}}_{p'}^{0 \dagger},
    = \sum_{p'} \bar{M}_{pp'}^{-1} \, \left( \bar{b}_{p'}^{0 \dagger} -
      \sum_{h} \bar{Z}_{p'h}^\ast \, b_h^{0 \dagger} \right), \label{ggop4}
  \end{align}
\end{subequations}
where the matrices $L$, $\bar{L}$, $M$, and $\bar{M}$ are here
determined by the solution to Eqs. \ref{lu1} and \ref{lu2}.

Because the transformation defined by Eqs. \ref{ggop1}--\ref{ggop4} is
canonical (we have explicitly ensured that anti-commutation rules are
preserved), we can invert the transformation using Eq. \ref{Tinv} as a
reference. We arrive at
\begin{subequations}
  \begin{align}
    b_h^{0 \dagger}
    &= \sum_{h'} \bar{L}_{h'h}^{\ast -1} \, b_{h'}^{1 \dagger} -
      \sum_{pp'} Z_{p'h} \, M_{pp'}^{\ast -1} \, \bar{b}_p^{1 \dagger},
    \label{ggop5} \\
    b_p^0
    &= \sum_{p'} \bar{M}_{p'p}^{-1} \, b_{p'}^1 + \sum_{hh'} Z_{ph'}
      \, L_{hh'}^{-1} \, \bar{b}_h^1,
    \label{ggop6} \\
    \bar{b}_h^0
    &= \sum_{h'} L_{h'h}^{-1} \, \bar{b}_{h'}^1 - \sum_{pp'}
      \bar{Z}_{p'h}^\ast \, \bar{M}_{pp'}^{-1} \, b_p^1,
    \label{ggop7} \\
    \bar{b}_p^{0 \dagger}
    &= \sum_{p'} M_{p'p}^{\ast -1} \, \bar{b}_{p'}^{1 \dagger} +
      \sum_{hh'} \bar{Z}_{ph'}^\ast \, \bar{L}_{hh'}^{\ast -1} b_h^{1
      \dagger}.
    \label{ggop8}
  \end{align}
\end{subequations}

We now use Eqs. \ref{ggop5}--\ref{ggop8} to write the global gradient
($\mathcal{G}$ and $\bar{\mathcal{G}}$) matrix elements in terms of
the local gradient ($G$ and $\bar{G}$) as
\begin{subequations}
  \label{glob_locgrad}
  \begin{align}
    \mathcal{G}_{ph}
    &= \sum_{p'h'} L_{h'h}^{\ast -1} \, M_{p'p}^{-1} \, G_{p'h'}
    = \left[ M^{\trans -1} \, G \, L^{\ast -1} \right]_{ph}, \\
    \bar{\mathcal{G}}_{ph}
    &= \sum_{p'h'} \bar{L}_{h'h}^{\ast -1} \, \bar{M}_{p'p}^{-1} \,
      \bar{G}_{p'h'}
    = \left[ \bar{M}^{\trans -1} \, \bar{G} \, \bar{L}^{\ast -1}
      \right]_{ph}.
  \end{align}
\end{subequations}

We close this subsection by noting that one has reached a solution to
the variational equations when the local gradient (and, consequently,
the global gradient) vanishes, {\em i.e.},
\begin{subequations}
  \begin{align}
    \frac{\partial}{\partial \, Z_{ph}^\ast} \, E &= 0, \\[4pt]
    \frac{\partial}{\partial \, \bar{Z}_{ph}^\ast} \, E &= 0.
  \end{align}
\end{subequations}

\section{Variational ansatz with projection operators \label{sec:proj_ansatz}}

We now turn our attention to states resulting from the action of
symmetry-restoring projection operators on symmetry-broken
determinants. We start by providing a brief description of the form of
the projection operators used. More details can be found in
Refs. \onlinecite{blaizot_ripka}, \onlinecite{ring_schuck}, or
\onlinecite{schmid2004}.

Consider a symmetry group $\hat{G}$, with elements $\{ \hat{g} \}$,
that commutes with the Hamiltonian. The group can be continuous or
discrete, but we shall assume for simplicity that it is Abelian. A
Slater determinant is symmetry broken if
\begin{equation}
  \hat{g} |\Phi \rangle \neq |\Phi \rangle,
\end{equation}
that is, if the determinant is not invariant upon action by the
elements $\{ \hat{g} \}$. The set of all $\left\{ \hat{g} |\Phi
\rangle \right\}$ is called the Goldstone manifold. The norm and the
matrix elements of commuting observables are the same within the
Goldstone manifold up to an arbitrary phase factor
\cite{blaizot_ripka}. It is well known \cite{peierls1957} that the
symmetry can be restored by diagonalization of the Hamiltonian among
the Goldstone manifold.

A projection operator can, in general, be written as
\begin{equation}
  \hat{P}^j = \frac{1}{L} \int_L d \theta \, w^j (\theta) \,
    \hat{R}_\theta,
  \label{projop}
\end{equation}
where $L$ is the volume of integration, $\hat{R}_\theta$ is an element
of the symmetry group in consideration, the index $j$ labels the
eigenvalue restored by means of the projection, and the coefficients
$w^j (\theta)$ correspond to the matrix elements of the operator
$\hat{R}_\theta$ among the irreducible representations of the
group. Evidently, for discrete groups the integration above is
replaced by a discrete sum. We shall drop the label $j$ henceforth for
simplicity of notation.

As an example of the projection operators discussed above, $S_z$
projection on a broken-symmetry determinant can be accomplished by
\begin{equation}
  \hat{P}^m = \frac{1}{4\pi} \int d \theta \, \exp \left[ i \theta
    \left( \hat{S}_z - m \right) \right],
\end{equation}
where an eigenfunction of $\hat{S}_z$ with eigenvalue $m$ is recovered
upon the action of the projection operator above.

We work with cases where $\hat{R}_\theta$ are single-particle rotation
operators that act on the HF ones according to
\begin{equation}
  b_k^\dagger (\theta) \equiv \hat{R}_\theta \, b_k^\dagger
    \hat{R}^{-1}_\theta = \sum_{j} D_{jk}^\ast \, \hat{R}_\theta \,
    c_j^\dagger \hat{R}^{-1}_\theta = \sum_{ij} R_{ij} (\theta) \,
    D_{jk}^\ast \, c_i^\dagger,
\end{equation}
where $R_{ij} (\theta) = \langle i | \hat{R}_\theta | j \rangle$ is
the matrix representation of $\hat{R}_\theta$ in the single-particle
basis.

We can now use the variational ansatz introduced in Eq. \ref{defPsi}
and put a projection operator in front of it. The proposed
wavefunction becomes
\begin{equation}
  \hat{P} |\Psi \rangle =
  \int d\theta \, w(\theta) \left[ c_1 \, \hat{R}_\theta |\Phi \rangle +
    c_2 \, \hat{R}_\theta |\overline{\Phi} \rangle \right].
  \label{defpPsi}
\end{equation}

The Hamiltonian expectation value of a wavefunction of the form of
Eq. \ref{defpPsi} can be written as
\begin{align}
  E [\Psi]
  &= \frac{\langle \Psi | \hat{P}^\dagger \, \hat{H} \, \hat{P} | \Psi \rangle}%
          {\langle \Psi | \hat{P}^\dagger \, \hat{P} | \Psi \rangle}
  = \frac{\langle \Psi | \hat{H} \, \hat{P} | \Psi \rangle}%
         {\langle \Psi | \hat{P} | \Psi \rangle} \nonumber \\[4pt]
  &= \int d\theta \, w(\theta)
  \sum_{\alpha,\beta=1}^2 y_{\alpha \beta} (\theta)
  \frac{\langle \Phi_\alpha | \hat{H} \, \hat{R}_\theta | \Phi_\beta \rangle}%
       {\langle \Phi_\alpha | \hat{R}_\theta | \Phi_\beta \rangle},
    \label{EpPsi} \\[4pt]
  y_{\alpha \beta} (\theta) &=
  \frac{c_\alpha^\ast \, c_\beta \langle \Phi_\alpha | \hat{R}_\theta
         | \Phi_\beta \rangle}%
       {\displaystyle \int d\theta \, w(\theta) \sum_{\alpha',\beta'=1}^2
         c_{\alpha'}^\ast \, c_{\beta'} \langle \Phi_{\alpha'} |
         \hat{R}_\theta | \Phi_{\beta'} \rangle}, \label{ypPsi}
\end{align}
where we have made the identifications $|\Phi_1 \rangle \equiv |\Phi
\rangle$ and $|\Phi_2 \rangle \equiv |\overline{\Phi} \rangle$. The
expressions for the matrix elements appearing in Eqs. \ref{EpPsi} and
\ref{ypPsi} are given in Appendix \ref{sec:app_matel}.

\subsection{Optimization of the projected ansatz $|\Psi \rangle$}

Our task is now to minimize the energy of our ansatz for the projected
state (Eq. \ref{EpPsi}) with respect to variations in the reference
determinants $|\Phi \rangle$ and $|\overline{\Phi} \rangle$. We will
closely follow the derivation we presented before (section
\ref{sec:opt}) for the optimization of the unprojected state.

The variation with respect to the coefficients $c_1$ and $c_2$ yields
a generalized eigenvalue problem similar to the one of Eqs. \ref{gev1}
and \ref{gev2}. In this case, $\mathbf{H}$ and $\mathbf{N}$ are $2
\times 2$ matrices given by
\begin{align}
  H_{\alpha \beta} &=
  \int d\theta \, w(\theta) \, \langle \Phi_\alpha | \hat{H} \,
    \hat{R}_\theta | \Phi_\beta \rangle, \\
  N_{\alpha \beta} &=
  \int d\theta \, w(\theta) \, \langle \Phi_\alpha | \hat{R}_\theta |
    \Phi_\beta \rangle.
\end{align}
Once again, only the lowest-energy solution is used in the variational
optimization.

The parametrization of the energy functional with respect to the
determinants $|\Phi \rangle$ and $|\overline{\Phi} \rangle$ is done in
the same way as it was done for the unprojected case
\cite{egido1995,schmid2004}. That is, we parametrize the energy
functional in terms of the Thouless' rotation matrices $Z$ and
$\bar{Z}$ acting upon $|\Phi \rangle$ and $|\overline{\Phi} \rangle$,
respectively.

The resulting local gradient is derived by using the definitions in
Eqs. \ref{LocalG1} and \ref{LocalG2}. We arrive at the expressions
\begin{subequations}
  \label{pgrad}
  \begin{align}
    G_{ph} &=
    \int d\theta \, w(\theta) \, \left\{
    - y_{11} (\theta) \, \frac{\langle \Phi | \bar{b}_h^\dagger \,
      \bar{b}_p \, \left( \hat{H} - E \right) \, \hat{R}_\theta | \Phi
      \rangle}{\langle \Phi | \hat{R}_\theta | \Phi \rangle}
    - y_{12} (\theta) \, \frac{\langle \Phi | \bar{b}_h^\dagger \,
      \bar{b}_p \, \left( \hat{H} - E \right) \, \hat{R}_\theta |
      \overline{\Phi} \rangle}{\langle \Phi | \hat{R}_\theta |
      \overline{\Phi} \rangle}
    \right\}, \\
    \bar{G}_{ph} &=
    \int d\theta \, w(\theta) \, \left\{
    - y_{21} (\theta) \, \frac{\langle \overline{\Phi} | b_h^\dagger \,
      b_p \, \left( \hat{H} - E \right) \, \hat{R}_\theta | \Phi
      \rangle}{\langle \overline{\Phi} | \hat{R}_\theta | \Phi \rangle}
    - y_{22} (\theta) \, \frac{\langle \overline{\Phi} | b_h^\dagger \,
      b_p \, \left( \hat{H} - E \right) \, \hat{R}_\theta |
      \overline{\Phi} \rangle}{\langle \overline{\Phi} | \hat{R}_\theta
      | \overline{\Phi} \rangle}
    \right\}.
  \end{align}
\end{subequations}
Here, $E$ is the energy corresponding to the state $|\Psi \rangle$
from Eq. \ref{defpPsi}. The explicit expressions for the matrix
elements appearing in Eq. \ref{pgrad} are given as part of Appendix
\ref{sec:app_matel}.  We finally note that the relationship between
the local gradient and the global gradient is the same as in the
unprojected case (see Eq. \ref{glob_locgrad}).

\section{Application to the one-dimensional Hubbard Hamiltonian \label{sec:hubbard}}

In this section we present the application of the ans\"atze discussed
previously to the one-dimensional Hubbard Hamiltonian \cite{essler}
with PBC. This describes a set of electrons in a lattice according to
\begin{equation}
  \hat{H} = -t \sum_{j, \sigma} \Big( c_{j, \sigma}^\dagger \, c_{j+1,
    \sigma} + c_{j+1, \sigma}^\dagger \, c_{j, \sigma} \Big) + U
    \sum_{j} c_{j, \uparrow}^\dagger \, c_{j, \uparrow} \, c_{j,
    \downarrow}^\dagger \, c_{j, \downarrow}.
  \label{hubbard}
\end{equation}
Here, $c_{j,\sigma}^\dagger$ creates an electron on site $j$ of the
lattice with $\sigma = \{ \uparrow, \downarrow \}$ $z$-projection of
spin. The first term in the Hamiltonian accounts for a negative ($t >
0$) kinetic energy that the electrons gain when they hop from one site
to a neighbor. The second term accounts for the ($U > 0$) repulsion
that opposite-spin electrons feel when they are in the same site. The
lattice used for this Hamiltonian is a finite one with $N_s$
sites. Periodic boundary conditions are assumed, which make the site
$N_s+k$ equivalent to the site $k$.

The 1D Hubbard Hamiltonian has been extensively studied, and our
purpose here is merely to test the flexibility that $N$-particle
Slater determinants constructed in terms of non-unitary canonical
transformations bring. With this in mind, ours should be regarded as a
proof of feasibility for calculations in finite many-fermion systems
based on non-unitary HF transformations. We point the interested
reader to the comprehensive book on the 1D Hubbard Hamiltonian by
Essler {\em et al}. \cite{essler}. Recent work on the 1D Hubbard
Hamiltonian with projected HF approximations has been done by Schmid
{\em et al}. \cite{schmid2005} and Tomita \cite{tomita2004}. We also
note that Lieb and Wu \cite{lieb1968} devised a set of equations from
which the exact eigenvalues of the 1D Hubbard Hamiltonian of
Eq. \ref{hubbard} can be obtained.

We have applied the methods described in the preceeding sections to
the 1D-Hubbard Hamiltonian. Our calculations have been performed with
an in-house code using the conjugate-gradient method described here
and in Ref. \onlinecite{egido1995} for the variational optimization of
HF-based states (see also Refs. \onlinecite{schmid2005} and
\onlinecite{rodriguez-guzman2012}). We have selected $U = 4\,t$ as a
representative on-site repulsion, corresponding to a strongly
correlated case ($U$ is of the order of the non-interacting
bandwidth). Nevertheless, our formalism can be used for any other $U$
value belonging to the weak, intermediate, or strong coupling
regimes. For all methods except the restricted HF (RHF), we have
constructed an initial guess of the HF transformation such that all
symmetries (spin, lattice momentum) are broken. This is sometimes
referred to as generalized HF (GHF) in the literature
\cite{jimenez-hoyos2011}. We have converged the HF states such that
the norm of the gradient is smaller than $10^{-5}$. For methods
involving S$_z$ projection we have chosen to recover states with
$\hat{S}_z$ eigenvalue $m=0$, as it is known that at half-filling the
ground state is always a singlet state \cite{lieb1989}. The exact
ground state energies, evaluated by solution to the Lieb--Wu equations
from Ref. \onlinecite{lieb1968}, have been obtained with an in-house
\verb+Mathematica+ notebook.

Table \ref{tab1} shows the total energies predicted by a variety of
methods for the ground state of the 1D Hubbard Hamiltonian at
half-filling ($N = N_s$, where $N$ is the number of electrons in the
system). It is evident from the results shown in Table \ref{tab1} that
nu-HF (defined by Eq. \ref{defPsi}), which uses the full flexibility
of a non-unitary HF transformation, is able to yield lower energies
than standard HF. This was expected, since it is at the very least a
two-configuration wavefunction. Similarly, nu-S$_z$HF (defined by
Eq. \ref{defpPsi}) yields lower energies than standard S$_z$-projected
HF.

\begin{table*}
  \caption{Total energies (in units of $t$, the hopping parameter) for
    the ground state of the $N$-site 1D Hubbard model Hamiltonian at
    half-filling with different approximate methods. We have set $U =
    4\,t$ for all calculations. \label{tab1}}

  \begin{ruledtabular}
    \begin{tabular}{r r r r r r r}
      $N$ &
      RHF\footnote{Restricted Hartree--Fock, {\em i.e.}, all
        symmetries of the Hamiltonian are preserved.} &
      HF\footnote{Symmetry-broken Hartree--Fock.} &
      nu-HF\footnote{Non-unitary Hartree--Fock, defined by
        Eq. \ref{defPsi}.} &
      S$_z$HF\footnote{S$_z$-projected Hartree--Fock (with $\hat{S}_z$
        eigenvalue $m=0$).} &
      nu-S$_z$HF\footnote{S$_z$-projected non-unitary Hartree--Fock,
        defined by Eq. \ref{defpPsi} (with $\hat{S}_z$ eigenvalue $m =
        0$).} &
      exact\footnote{Obtained by solution to the Lieb--Wu equations of
        Ref. \onlinecite{lieb1968}.} \\
      \hline
        8  &   -1.656\,854  &   -3.748\,562  &   -3.969\,123  &   -4.163\,645  &   -4.342\,058  &   -4.603\,5 \\  
       12  &   -2.928\,203  &   -5.629\,064  &   -5.848\,959  &   -6.068\,077  &   -6.316\,985  &   -6.920\,4 \\  
       16  &   -4.109\,358  &   -7.505\,674  &   -7.722\,392  &   -7.948\,679  &   -8.231\,962  &   -9.214\,4 \\  
       24  &   -6.383\,016  &  -11.258\,526  &  -11.472\,354  &  -11.703\,719  &  -12.011\,295  &  -13.795\,8 \\  
       32  &   -8.612\,682  &  -15.011\,368  &  -15.224\,875  &  -15.457\,467  &  -15.777\,256  &  -18.379\,4 \\  
       48  &  -13.028\,207  &  -22.517\,052  &  -22.730\,518  &  -22.963\,973  &  -23.291\,156  &  -27.552\,4 \\  
       64  &  -17.421\,870  &  -30.022\,735  &  -30.236\,201  &  -30.470\,037  &  -30.798\,938  &  -36.728\,7 \\  
       96  &  -26.187\,360  &  -45.034\,103  &  -45.247\,569  &  -45.481\,765  &  -45.811\,121  &  -55.084\,7 \\  
      128  &  -34.941\,935  &  -60.045\,471  &  -60.258\,936  &  -60.493\,305  &  -60.822\,554  &  -73.442\,4 \\  
      192  &  -52.440\,176  &  -90.068\,206  &  -90.281\,672  &  -90.516\,208  &  -90.845\,293  & -110.159\,4 \\  
      256  &  -69.932\,961  & -120.090\,941  & -120.304\,407  & -120.539\,025  & -120.868\,029  & -146.877\,2 \\  
    \end{tabular}
  \end{ruledtabular}
\end{table*}

\begin{figure*}
  \includegraphics[scale=1.0]{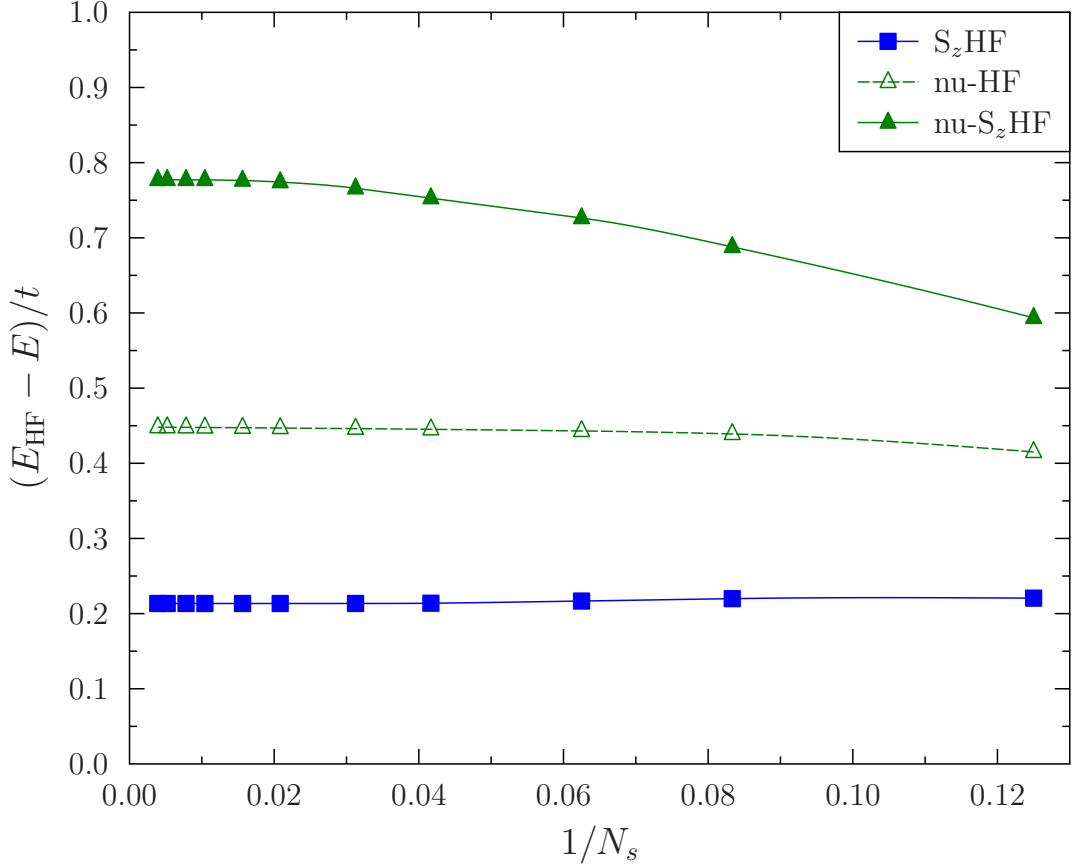}
  \caption{Total correlation energy (in units of $t$), predicted by
    S$_z$HF and the non-unitary based ans\"atze, for 1D Hubbard model
    calculations as a function of the number of sites $N_s$. The
    calculations were performed at half-filling, with $U = 4\,t$. The
    correlation energy has been defined with respect to the
    broken-symmetry HF solution. \label{fig:corr}}
\end{figure*}

It is less evident that the total correlation energy, defined
here \footnote{Note that this definition is not the one suggested by
  Lowdin \cite{lowdin1955} and commonly used in quantum chemistry,
  in which the correlation energy is defined with respect to the
  symmetry-preserving RHF solution.} as the difference with respect to
the energy of the broken-symmetry HF solution, should tend to a
non-zero constant with increasing lattice size. This is the case, as
shown in Fig. \ref{fig:corr}. In any case, the correlation energy per
particle predicted by all approximate methods considered in Table
\ref{tab1} goes to zero as $N \to \infty$. This is a reflection of the
limited flexibility that the projected HF and the non-unitary based
ans\"atze still have. Note, however, that even if the energy per
particle becomes the same as $N \to \infty$, the total energy and the
wavefunction itself are different from the symmetry broken HF
solution.

It is interesting to observe that simple ans\"atze such as nu-HF or
nu-S$_z$HF can be useful to describe finite-size lattices where they
can capture a significant part of the correlation.  For $N=12$, for
which the exact ground state energy is $-6.9204\,t$, nu-HF recovers
$17\,\%$, S$_z$HF recovers $34\,\%$, and nu-S$_z$HF recovers $53\,\%$
of the missing correlation energy in the broken-symmetry HF
solution. Full spin and linear momentum projection may be used to
recover even a larger fraction of correlation energy, as has been
shown for projected HF methods in small size Hubbard 1D or 2D lattices
\cite{schmid2005,rodriguez-guzman2012}.

\subsection*{Comparison with other two-determinant approaches}

The results shown so far indicate that nu-HF and nu-S$_z$HF improve
upon HF and S$_z$HF, respectively. This is due to a combination of the
more general canonical transformation being used and the fact that
nu-HF and nu-S$_z$HF are explicitly constructed as two-determinant
configurations.

It is interesting to compare the non-unitary based ans\"atze discussed
in this paper with other two-determinant ans\"atze resulting from a
{\em single, unitary canonical transformation}. We have already
discussed that more general two-determinant ans\"atze, where each
configuration results from an independent HF-transformation, have the
same flexibility as the non-unitary approaches considered in this
work, something we have verified numerically.

One can think of several ways to construct a two-determinant ansatz
based on a single, unitary canonical transformation. Our experience
shows that symmetry-projection approaches are very effective in
capturing electron correlations. In this sense, several two-element
symmetry groups can be used in the 1D periodic Hubbard Hamiltonian to
build a two-state Goldstone manifold: the complex-conjugation group
built with the elements $\{ \hat{I}, \hat{K} \}$, where $\hat{I}$ is
the identity operator and $\hat{K}$ is the complex conjugation
operator, the time-reversal group built with the elements $\{ \hat{I},
\hat{\Theta} \}$, where $\hat{\Theta} = \exp (i\,\pi\,\hat{S}_y) \,
\hat{K}$ is the time-reversal operator, or the $C_2$ group for even
lattices built with the elements $\{ \hat{I}, \hat{C}_{N_s/2} \}$,
where $\hat{C}_{N_s/2}$ is the operator performing a 180-degree
rotation of the lattice. We here consider the complex conjugation
group as a representative example. In this subsection, we compare our
non-unitary based ans\"atze with KHF, or complex-conjugation restored
HF, and KS$_z$HF, or complex-conjugation and S$_z$-projected HF:
\begin{align}
  |\Psi^{\mathrm{KHF}} \rangle &= c_1 |\Phi \rangle + c_2 \, \hat{K}
    |\Phi \rangle, \\
  |\Psi^{\mathrm{KS}_z\mathrm{HF}} \rangle &= c_1 \, \hat{P}^{S_z}
    |\Phi \rangle + c_2 \, \hat{P}^{S_z} \, \hat{K} |\Phi \rangle,
\end{align}
where $| \Phi \rangle$ is an $N$-particle Slater determinant and
$\hat{P}^{S_z}$ is the $S_z$ projection operator (onto $m = 0$).

\begin{figure*}
  \includegraphics[scale=1.0]{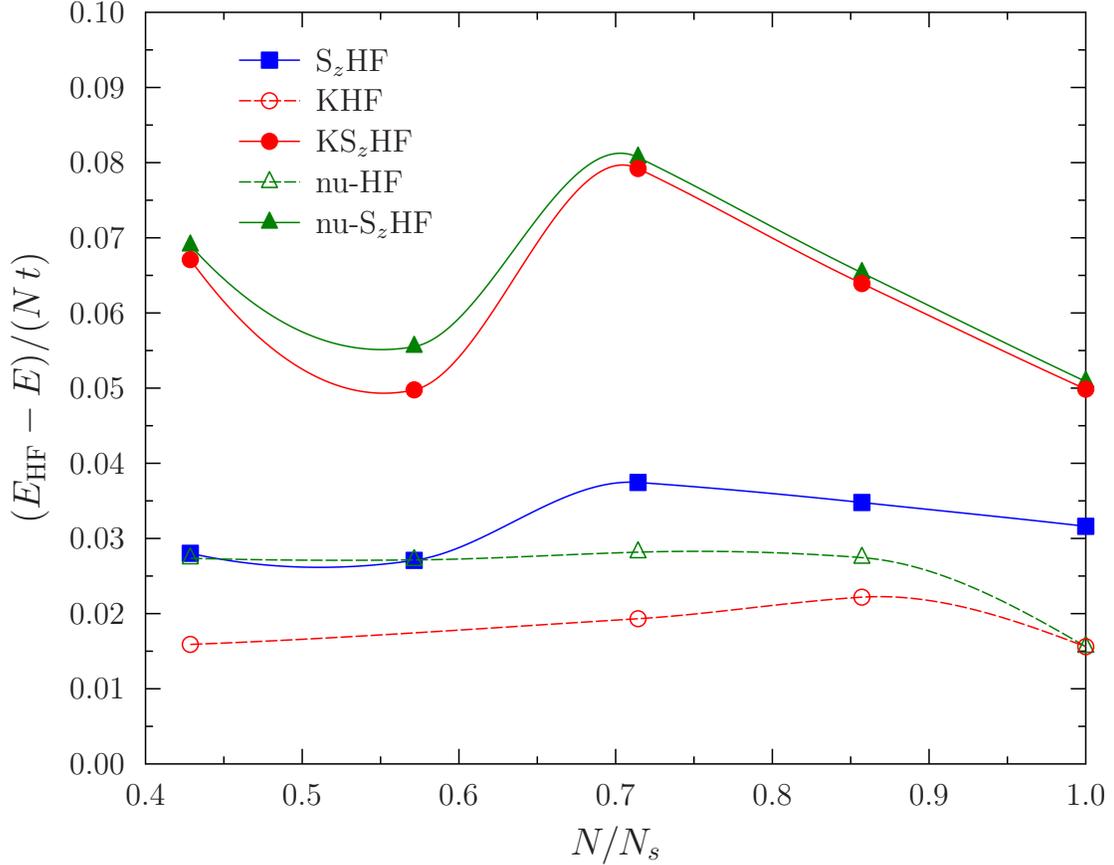}
  \caption{Correlation energy per electron (in units of $t$),
    predicted by a variety of approximate methods for a 14-site
    periodic 1D Hubbard model as a function of $N/N_s$. The
    correlation energy has been defined with respect to the
    broken-symmetry HF solution. We were unable to converge KHF for
    $N=8$. \label{fig:corr2}}
\end{figure*}

Figure \ref{fig:corr2} shows the correlation energy per electron
predicted by a variety of approximate methods for a 14-site periodic
1D Hubbard model as a function of the hole-filling ($N/N_s$). It is
interesting to note that at half-filling ($N/N_s = 1$) nu-HF and KHF
yield exactly the same correlation energy. In this sense, the full
flexibility of the non-unitary transformation is not being exploited
in the solution. At other fillings, on the other hand, nu-HF is able
to improve substantially over KHF. In contrast, nu-S$_z$HF yields
lower energies (or larger correlation energies) at all fillings, even
though the improvement is only marginal in some cases.

Overall, there is no guarantee that introducing more flexibility into
an approximate wavefunction will result in lower energies for every
system. We have shown, however, that ans\"atze based on a non-unitary
canonical transformation yield lower energies than HF or projected-HF
methods. They even yield lower energies than KHF or projected-KHF
solutions in some cases, despite the fact that complex-conjugation
projected wavefunctions are also two-determinant configurations, even
if they result from a single, {\em unitary} canonical transformation.

\section{Conclusions}

The HF and the HFB wavefunctions constitute the building blocks upon
which more elaborate many-body methods rely. They are built out of a
set of independent quasi-particles resulting from a linear unitary
canonical transformation of elementary fermion operators. In this
work, we have explored the possibility of relaxing the unitarity
condition within a HF-type formalism in order to have more variational
flexibility in the considered wavefunctions.

The properties of $N$-particle Slater determinants constructed from a
set of HF-type operators resulting from a non-unitary canonical
transformation of fermion operators have been discussed. We have
derived the corresponding Thouless' theorem for such states, which
allowed us to compute matrix elements in an efficient way by
application of Wick's theorem \cite{blaizot_ripka}.

An ansatz based on a single Slater determinant is incapable of
utilizing the full flexibility of a non-unitary transformation. We
have therefore introduced a two-determinant ansatz, defined by
Eq. \ref{defPsi}, where all the degrees of freedom of a HF-type
non-unitary transformation are used. This, however, is not a
limitation of the non-unitary transformation. One could work with
other more general ans\"atze used in many-body theory that utilize an
$N$-particle Slater determinant as a starting point.

Symmetry-breaking is commonly used within a HF formalism to access
relevant correlations that are otherwise difficult to obtain starting
from a symmetry-preserving Slater determinant. In this sense, a
non-unitary transformation provides additional degrees of freedom that
can be used in the variational problem. A symmetry-broken wavefunction
is, nevertheless, still unphysical; we advocate the use of projection
techniques out of a symmetry-broken intrinsic state, within a
variation-after-projection approach, to access the relevant
correlations resulting from large quantum fluctuations. This can be
done, as we have shown in the present work, in combination with a
non-unitary canonical transformation, affording even more flexibility
than that which a projected HF state based on a unitary HF
transformation has. A non-unitary based projected HF scheme aims to
provide an accurate description of a many-particle system with a
limited number of configurations, still a far-reaching problem in
fields such as nuclear and condensed matter physics as well as in
quantum chemistry.

Finally, we note that our formalism can also be used in the
optimization of $N$-particle Slater determinants that are considered
as approximations to the left- and right-eigenvectors of non-Hermitian
Hamiltonians. In particular, our work can be directly applied to
non-Hermitian Hamiltonians with real eigenvalues, such as those
resulting from similarity transformations of a standard Hermitian one.

The extension of this work to the full non-unitary Bogoliubov
transformation is possible and will be presented in a forthcoming
publication \cite{jimenez-hoyos2012b}.

\section*{Acknowledgments}

This work is supported by the US Department of Energy, Office of Basic
Energy Sciences, (DE-FG02-09ER16053), the National Science Foundation
(CHE-1110884), and the Welch Foundation (C-0036).

\appendix

\section{Proof of Thouless' thorem \label{sec:app_thou}}

In order to prove the extension to Thouless' theorem stated in section
\ref{sec:Thou}, we start by introducing the operators $\{
b_h^{\dagger}, b_p, \bar{b}_h, \bar{b}_p^\dagger \}$ and $\{
d_h^\dagger, d_p, \bar{d}_h, \bar{d}_p^\dagger \}$, such that $\{
b_h^\dagger, b_p \}$ kill the vacuum $|\Phi_0 \rangle$ and $\{
\bar{b}_h^\dagger, \bar{b}_p \}$ kill the vacuum $|\overline{\Phi}_0
\rangle$, while $\{ d_h^\dagger, d_p \}$ annihilate the vacuum
$|\Phi_1 \rangle$ and $\{ \bar{d}_h^\dagger, \bar{d}_p \}$ annihilate
the vacuum $|\overline{\Phi}_1 \rangle$.  We assume that both sets
obey the anti-commutation rules defined by Eq. \ref{HFanticom}. We
explicitly write these operators in the form of Eqs. \ref{operator1}
and \ref{operator2}; that is,
\begin{center}
  \begin{tabular}{c @{\hspace{1cm}} c}
    $b_h^\dagger = \displaystyle \sum_j D^{0 \ast}_{jh} \,
      c_j^\dagger$, &
    $ b_p = \displaystyle \sum_j D^{0}_{jp} \, c_j$ \\
    $\bar{b}_h = \displaystyle \sum_j \bar{D}^{0}_{jh} \, c_j$, &
    $\bar{b}_p^\dagger = \displaystyle \sum_j \bar{D}^{0 \ast}_{jp}
      \, c_j^\dagger$, \\
    $d_h^\dagger = \displaystyle \sum_j D^{1 \ast}_{jh} \,
      c_j^\dagger$, &
    $d_p = \displaystyle \sum_j D^{1}_{jp} \, c_j$, \\
    $\bar{d}_h = \displaystyle \sum_j \bar{D}^{1}_{jh} \, c_j$, &
    $\bar{d}_p^\dagger = \displaystyle \sum_j \bar{D}^{1 \ast}_{jp}
      \, c_j^\dagger$,
  \end{tabular}
\end{center}
where the superscripts on the matrices indicate the state to which the
operators correspond.

We can now relate the operators $\{ d_h^\dagger, d_p \}$ to the
operators $\{ b_h^{\dagger}, b_p \}$ by using the inverse
transformation discussed in Eq. \ref{HFinv}. We arrive at
\begin{subequations}
  \label{defd}
  \begin{align}
    d_h^\dagger &=
    \sum_{h'} L_{h'h}^\ast \, b_{h'}^\dagger + \sum_{p} Y_{ph}^\ast
      \, \bar{b}_p^\dagger, \\
    d_p &=
    \sum_{p'} M_{p'p} \, b_{p'} + \sum_{h} \tilde{Y}_{hp} \,
      \bar{b}_h,
  \end{align}
\end{subequations}
where we have set
\begin{subequations}
  \label{defLMY}
  \begin{align}
    L_{h'h} &= \left( \bar{D}^{0 \dagger} \, D^1 \right)_{h'h},
      \label{defL}\\
    M_{p'p} &= \left( \bar{D}^{0 \dagger} \, D^1 \right)_{p'p}, \\
    Y_{ph} &= \left( D^{0 \dagger} \, D^1 \right)_{ph}, \\
    \tilde{Y}_{hp} &= \left( D^{0 \dagger} \, D^1 \right)_{hp}.
  \end{align}
\end{subequations}

We now assume that the $N \times N$ matrix $L$ is invertible, which is
only true if $\langle \overline{\Phi}_0 | \Phi_1 \rangle \neq 0$ (see
Eq. \ref{overlap}). In such a case, the matrix $M$ is also
invertible. We now introduce the operators
\begin{subequations}
  \label{defdtilde}
  \begin{align}
    \tilde{d}_h^\dagger &= \displaystyle \sum_{h'} \left( L^{\ast -1}
      \right)_{h'h} \, d_{h'}^{\dagger}, \\
    \tilde{d}_p &= \displaystyle \sum_{p'} \left( M^{-1} \right)_{p'p}
      \, d_{p'}.
  \end{align}
\end{subequations}

Inserting Eq. \ref{defdtilde} into Eq. \ref{defd}, we arrive at
\begin{subequations}
  \begin{align}
    \tilde{d}_h^\dagger &= b_h^\dagger + \sum_{p} Z_{ph} \,
      \bar{b}_p^\dagger, \\
    \tilde{d}_p &= b_p + \sum_{h} W_{ph} \, \bar{b}_h,
  \end{align}
\end{subequations}
where we have set
\begin{subequations}
  \label{defZ}
  \begin{align}
    Z_{ph} &= \displaystyle \sum_{h'} Y_{ph'}^\ast \left( L^{\ast -1}
      \right)_{h'h}, \\
    W_{ph} &= \displaystyle \sum_{p'} \tilde{Y}_{hp'} \left( M^{-1}
      \right)_{p'p}.
  \end{align}
\end{subequations}
In fact, by computing the anti-commutation rules among the operators
$\{ \tilde{d}_h^\dagger, \tilde{d}_p \}$, one can readily conclude
that $W = -Z$. This also implies that if $L$ is invertible, then so is
$M$. The transformed operators become
\begin{subequations}
  \begin{align}
    \tilde{d}_h^\dagger &= b_h^\dagger + \sum_{p} Z_{ph} \,
      \bar{b}_p^\dagger, \label{th1} \\
    \tilde{d}_p &= b_p - \sum_{h} Z_{ph} \, \bar{b}_h. \label{th2}
  \end{align}
\end{subequations}

We are now in a position to investigate whether the transformed
operators, defined by Eqs. \ref{th1} and \ref{th2}, annihilate the
vacuum defined by Eq. \ref{Thou2}. We start by evaluating the
commutators
\begin{subequations}
  \begin{align}
    \left[ b_h^\dagger, \exp \left( \sum_{p'h'} Z_{p'h'} \,
      \bar{b}_{p'}^\dagger \, \bar{b}_{h'} \right) \right] &=
    \left( -\sum_p Z_{ph} \bar{b}_p^\dagger \right) \exp \left(
      \sum_{p'h'} Z_{p'h'} \, \bar{b}_{p'}^\dagger \, \bar{b}_{h'}
      \right), \\[4pt]
    \left[ b_p, \exp \left( \sum_{p'h'} Z_{p'h'} \,
      \bar{b}_{p'}^\dagger \, \bar{b}_{h'} \right) \right] &=
    \left( \sum_h Z_{ph} \bar{b}_h \right) \exp \left( \sum_{p'h'}
      Z_{p'h'} \, \bar{b}_{p'}^\dagger \, \bar{b}_{h'} \right).
  \end{align}
\end{subequations}
The operators from Eqs. \ref{th1} and \ref{th2} act on the vacuum of
Eq. \ref{Thou2} as
\begin{subequations}
  \begin{align}
    \tilde{d}_h^\dagger \exp \left( \sum_{ph} Z_{ph} \,
      \bar{b}_p^\dagger \, \bar{b}_h \right) |\Phi_0 \rangle &=
    \left( -\sum_p Z_{ph} \bar{b}_p^\dagger + \sum_p Z_{ph}
      \bar{b}_p^\dagger \right) \exp \left( \sum_{p'h'} Z_{p'h'} \,
      \bar{b}_{p'}^\dagger \, \bar{b}_{h'} \right) |\Phi_0 \rangle
    = 0, \\[4pt]
    \tilde{d}_p \exp \left( \sum_{ph} Z_{ph} \,
      \bar{b}_p^\dagger \, \bar{b}_h \right) |\Phi_0 \rangle &=
    \left( \sum_h Z_{ph} \bar{b}_h - \sum_h Z_{ph}
      \bar{b}_h \right) \exp \left( \sum_{p'h'} Z_{p'h'} \,
      \bar{b}_{p'}^\dagger \, \bar{b}_{h'} \right) |\Phi_0 \rangle
    = 0.
  \end{align}
\end{subequations}
This essentially completes the proof. $\{ \tilde{d}_h^\dagger,
\tilde{d}_p \}$ annihilate the r.h.s. of Eq. \ref{th2}. The operators
$\{ d_h^\dagger, d_p \}$ that kill the vacuum $|\Phi_1 \rangle$ on
the l.h.s. of Eq. \ref{th2} are simple linear combinations of $\{
\tilde{d}_h^\dagger, \tilde{d}_p \}$; $N$-particle Slater determinants
built from either sets of operators are the same up to a normalization
factor.

\section{Matrix elements appearing in projected states \label{sec:app_matel}}

Here, we provide explicit formulas for the matrix elements appearing
in the energy expression and in the local gradient from the
variational ansatz based on projected states.

The overlap kernels appearing in Eq. \ref{ypPsi} are evaluated as
\begin{subequations}
  \begin{align}
    \langle \Phi | \hat{R}_\theta | \Phi \rangle &=
      \mathrm{det}_N \, D^\trans \, R (\theta) \, D^\ast, \\
    \langle \overline{\Phi} | \hat{R}_\theta | \Phi \rangle &=
      \mathrm{det}_N \, \bar{D}^\trans \, R (\theta) \, D^\ast, \\
    \langle \Phi | \hat{R}_\theta | \overline{\Phi} \rangle &=
      \mathrm{det}_N \, D^\trans \, R (\theta) \, \bar{D}^\ast, \\
    \langle \overline{\Phi} | \hat{R}_\theta | \overline{\Phi} \rangle &=
      \mathrm{det}_N \, \bar{D}^\trans \, R (\theta) \, \bar{D}^\ast.
  \end{align}
\end{subequations}

The Hamiltonian kernels appearing in Eq. \ref{EpPsi} are evaluated in
terms of transition density matrices as
\begin{align}
  \frac{\langle \Phi_\alpha | \hat{H} \, \hat{R}_\theta | \Phi_\beta \rangle}%
       {\langle \Phi_\alpha | \hat{R}_\theta | \Phi_\beta \rangle} &=
  \mathrm{Tr} \left( h \, \rho^{\alpha \beta} (\theta) + \frac{1}{2}
    \, \Gamma^{\alpha \beta} (\theta) \, \rho^{\alpha \beta} (\theta)
    \right), \\
  \Gamma_{ik}^{\alpha \beta} (\theta) &=
  \sum_{jl} \langle ij | \hat{v} | kl \rangle \, \rho_{lj}^{\alpha
    \beta} (\theta).
\end{align}

The transition density matrices are in turn given by
\begin{subequations}
  \begin{align}
    \rho_{ki}^{11} (\theta) &=
    \frac{\langle \Phi | c_i^\dagger \, c_k \, \hat{R}_\theta | \Phi \rangle}%
         {\langle \Phi | \hat{R}_\theta | \Phi \rangle}
    = \sum_h D_{ih} \, \bar{D}_{kh}^\ast + \sum_{ph} D_{ih} \,
      \mathcal{Z}^{(11)}_{ph} (\theta) \, D_{kp}^\ast, \\[4pt]
    \rho_{ki}^{12} (\theta) &=
    \frac{\langle \Phi | c_i^\dagger \, c_k \, \hat{R}_\theta | \overline{\Phi} \rangle}%
         {\langle \Phi | \hat{R}_\theta | \overline{\Phi} \rangle}
    = \sum_h D_{ih} \, \bar{D}_{kh}^\ast + \sum_{ph} D_{ih} \,
      \mathcal{Z}^{(12)}_{ph} (\theta) \, D_{kp}^\ast, \\[4pt]
    \rho_{ki}^{21} (\theta) &=
    \frac{\langle \overline{\Phi} | c_i^\dagger \, c_k \, \hat{R}_\theta | \Phi \rangle}%
         {\langle \overline{\Phi} | \hat{R}_\theta | \Phi \rangle}
    = \sum_h \bar{D}_{ih} \, D_{kh}^\ast + \sum_{ph} \bar{D}_{ih} \,
      \mathcal{Z}^{(21)}_{ph} (\theta) \, \bar{D}_{kp}^\ast, \\[4pt]
    \rho_{ki}^{22} (\theta) &=
    \frac{\langle \overline{\Phi} | c_i^\dagger \, c_k \, \hat{R}_\theta | \overline{\Phi} \rangle}%
         {\langle \overline{\Phi} | \hat{R}_\theta | \overline{\Phi} \rangle}
    = \sum_h \bar{D}_{ih} \, D_{kh}^\ast + \sum_{ph} \bar{D}_{ih} \,
      \mathcal{Z}^{(22)}_{ph} (\theta) \, \bar{D}_{kp}^\ast.
  \end{align}
\end{subequations}
Here,
\begin{subequations}
  \begin{align}
    \mathcal{Z}^{(11)}_{ph} (\theta) &=
    \sum_{h'} \left( \bar{D}^\trans \, R (\theta) \, D^\ast \right)_{ph'}
      \left( \mathcal{L}^{(11) \ast -1} (\theta) \right)_{h'h}, \\
    \mathcal{Z}^{(12)}_{ph} (\theta) &=
    \sum_{h'} \left( \bar{D}^\trans \, R (\theta) \, \bar{D}^\ast \right)_{ph'}
      \left( \mathcal{L}^{(12) \ast -1} (\theta) \right)_{h'h}, \\
    \mathcal{Z}^{(21)}_{ph} (\theta) &=
    \sum_{h'} \left( D^\trans \, R (\theta) \, D^\ast \right)_{ph'}
      \left( \mathcal{L}^{(21) \ast -1} (\theta) \right)_{h'h}, \\
    \mathcal{Z}^{(22)}_{ph} (\theta) &=
    \sum_{h'} \left( D^\trans \, R (\theta) \, \bar{D}^\ast \right)_{ph'}
      \left( \mathcal{L}^{(22) \ast -1} (\theta) \right)_{h'h},
  \end{align}
\end{subequations}
and
\begin{subequations}
  \begin{align}
    \mathcal{L}^{(11)}_{h'h} (\theta) &=
    \left( D^\dagger \, R^\ast (\theta) \, D \right)_{h'h}, \\
    \mathcal{L}^{(12)}_{h'h} (\theta) &=
    \left( D^\dagger \, R^\ast (\theta) \, \bar{D} \right)_{h'h}, \\
    \mathcal{L}^{(21)}_{h'h} (\theta) &=
    \left( \bar{D}^\dagger \, R^\ast (\theta) \, D \right)_{h'h}, \\
    \mathcal{L}^{(22)}_{h'h} (\theta) &=
    \left( \bar{D}^\dagger \, R^\ast (\theta) \, \bar{D} \right)_{h'h}.
  \end{align}
\end{subequations}

The overlap-like matrix elements appearing in the local gradient
(Eq. \ref{pgrad}) can be evaluated as
\begin{subequations}
  \begin{align}
    \frac{\langle \Phi | \bar{b}_h^\dagger \, \bar{b}_p \,
           \hat{R}_\theta | \Phi \rangle}%
         {\langle \Phi | \hat{R}_\theta | \Phi \rangle} &=
    \sum_{mn} \bar{D}_{mh}^\ast \, \bar{D}_{np} \, \rho^{11}_{nm} (\theta),
    \\[4pt]
    \frac{\langle \overline{\Phi} | b_h^\dagger \, b_p \,
           \hat{R}_\theta | \Phi \rangle}%
         {\langle \overline{\Phi} | \hat{R}_\theta | \Phi \rangle} &=
    \sum_{mn} D_{mh}^\ast \, D_{np} \, \rho^{21}_{nm} (\theta),
    \\[4pt]
    \frac{\langle \Phi | \bar{b}_h^\dagger \, \bar{b}_p \,
           \hat{R}_\theta | \overline{\Phi} \rangle}%
         {\langle \Phi | \hat{R}_\theta | \overline{\Phi} \rangle} &=
    \sum_{mn} \bar{D}_{mh}^\ast \, \bar{D}_{np} \, \rho^{12}_{nm} (\theta),
    \\[4pt]
    \frac{\langle \overline{\Phi} | b_h^\dagger \, b_p \,
           \hat{R}_\theta | \overline{\Phi} \rangle}%
         {\langle \overline{\Phi} | \hat{R}_\theta | \overline{\Phi} \rangle} &=
    \sum_{mn} D_{mh}^\ast \, D_{np} \, \rho^{22}_{nm} (\theta).
  \end{align}
\end{subequations}

Similarly, the Hamiltonian-like matrix elements in Eq. \ref{pgrad} can
be evaluated as
\begin{subequations}
  \begin{align}
    \frac{\langle \Phi | \bar{b}_h^\dagger \, \bar{b}_p \, \hat{H} \,
            \hat{R}_\theta | \Phi \rangle}%
         {\langle \Phi | \hat{R}_\theta | \Phi \rangle} &=
    \sum_{mn} \bar{D}_{mh}^\ast \, \bar{D}_{np} \, \rho^{11}_{nm} (\theta) \,
    \frac{\langle \Phi | \hat{H} \, \hat{R}_\theta | \Phi \rangle}%
         {\langle \Phi | \hat{R}_\theta | \Phi \rangle} \nonumber \\ &+
    \sum_{mn} \sum_{ik} \bar{D}_{mh}^\ast \, \bar{D}_{np} \, \left(
      h_{ik} + \Gamma^{11}_{ik} (\theta) \right) \, \rho^{11}_{km} (\theta)
      \, \left( \delta_{ni} - \rho^{11}_{ni} (\theta) \right),
    \\[4pt]
    \frac{\langle \overline{\Phi} | b_h^\dagger \, b_p \, \hat{H} \, 
            \hat{R}_\theta | \Phi \rangle}%
         {\langle \overline{\Phi} | \hat{R}_\theta | \Phi \rangle} &=
    \sum_{mn} D_{mh}^\ast \, D_{np} \, \rho^{21}_{nm} (\theta) \,
    \frac{\langle \overline{\Phi} | \hat{H} \, \hat{R}_\theta | \Phi \rangle}%
         {\langle \overline{\Phi} | \hat{R}_\theta | \Phi \rangle} \nonumber \\ &+
    \sum_{mn} \sum_{ik} D_{mh}^\ast \, D_{np} \, \left( h_{ik} +
      \Gamma^{21}_{ik} (\theta) \right) \, \rho^{21}_{km} (\theta) \,
      \left( \delta_{ni} - \rho^{21}_{ni} (\theta) \right),
    \\[4pt]
    \frac{\langle \Phi | \bar{b}_h^\dagger \, \bar{b}_p \, \hat{H} \,
            \hat{R}_\theta | \overline{\Phi} \rangle}%
         {\langle \Phi | \hat{R}_\theta | \overline{\Phi} \rangle} &=
    \sum_{mn} \bar{D}_{mh}^\ast \, \bar{D}_{np} \, \rho^{12}_{nm} (\theta) \,
    \frac{\langle \Phi | \hat{H} \, \hat{R}_\theta | \overline{\Phi} \rangle}%
         {\langle \Phi | \hat{R}_\theta | \overline{\Phi} \rangle} \nonumber \\ &+
    \sum_{mn} \sum_{ik} \bar{D}_{mh}^\ast \, \bar{D}_{np} \, \left(
      h_{ik} + \Gamma^{12}_{ik} (\theta) \right) \, \rho^{12}_{km} (\theta)
      \, \left( \delta_{ni} - \rho^{12}_{ni} (\theta) \right),
    \\[4pt]
    \frac{\langle \overline{\Phi} | b_h^\dagger \, b_p \, \hat{H} \, 
            \hat{R}_\theta | \overline{\Phi} \rangle}%
         {\langle \overline{\Phi} | \hat{R}_\theta | \overline{\Phi} \rangle} &=
    \sum_{mn} D_{mh}^\ast \, D_{np} \, \rho^{22}_{nm} (\theta) \,
    \frac{\langle \overline{\Phi} | \hat{H} \, \hat{R}_\theta | \overline{\Phi} \rangle}%
         {\langle \overline{\Phi} | \hat{R}_\theta | \overline{\Phi} \rangle} \nonumber \\ &+
    \sum_{mn} \sum_{ik} D_{mh}^\ast \, D_{np} \, \left( h_{ik} +
      \Gamma^{22}_{ik} (\theta) \right) \, \rho^{22}_{km} (\theta) \,
      \left( \delta_{ni} - \rho^{22}_{ni} (\theta) \right).
  \end{align}
\end{subequations}

\bibliography{manuscript}

\end{document}